\documentclass[prd,aps,superscriptaddress,10pt,nobibnotes,notitlepage,twocolumn,longbibliography,nofootinbib,preprintnumbers]{revtex4-1}

\usepackage[normalem]{ulem}
\usepackage{sidecap}
\usepackage[latin1]{inputenc}
\usepackage{graphicx}
\usepackage{float}
\usepackage{latexsym}
\usepackage{amssymb}
\usepackage{amsmath}
\usepackage{amsfonts}
\usepackage{ragged2e}

\usepackage[colorlinks=true,citecolor=blue,hyperfootnotes=false]{hyperref}

\usepackage{cool}
\usepackage{dcolumn}
\usepackage{textcomp}
\usepackage[dvipsnames]{xcolor}
\usepackage{xfrac}
\usepackage{slashed}
\usepackage{multirow}

\def\be{\begin{equation}}
\def\ee{\end{equation}}
\def\bea{\begin{eqnarray}}
\def\eea{\end{eqnarray}}
\def\bg{\begin{eqnarray}}
\def\nd{\end{eqnarray}}
\def\beq{\begin{equation}}
\def\eeq{\end{equation}}

\begin{document}

\preprint{KCL-PH-TH/2026-09} 

\title{Spontaneous Baryogenesis and Primordial Black Hole Dark Matter from Ultra-Slow-Roll Inflation}

\author{Shyam Balaji}
\email{shyam.balaji@kcl.ac.uk}
\affiliation{Physics Department, King's College London, Strand, London, WC2R 2LS, United Kingdom}

\begin{abstract}
We propose a unified framework where the totality of dark matter (DM), the baryon asymmetry of the universe, and a detectable stochastic gravitational wave (GW) background originate from ultra-slow-roll (USR) inflation. The drastic suppression of the inflaton velocity during the USR phase, required for primordial black hole (PBH) DM production, can also set the initial conditions for spontaneous baryogenesis via a derivative coupling. This mechanism establishes a predictive correlation between the PBH abundance and the baryon yield, effectively fixing the reheating temperature $T_\textrm{reh}$ as a function of the post-peak spectral slope of the primordial power spectrum and the tensor-to-scalar ratio on CMB scales $r_\textrm{CMB}$. We perform a simple scan of the parameter space, demonstrating that while ``flat'' spectral tails allow for high-scale inflation ($r_{\rm CMB} \lesssim 10^{-3}$, $T_{\rm reh} \lesssim 10^{14} \text{ GeV}$) with a small wedge of tensor-to-scalar ratios potentially accessible to future CMB B-mode experiments, steep spectral tails enforce drastically lower scale inflation with an unobservably small $r_{\rm CMB}$ to avoid baryon overproduction. This degeneracy can be broken by GW astronomy: while LISA and DECIGO are capable of detecting the induced GW background associated with asteroid-mass PBH DM, the Einstein Telescope (ET) can act as a spectral discriminator, sensitive only to the broadband signals of high-scale scenarios.
\end{abstract}
\maketitle

%%%%%%%%%%%%%%%%%%%%%%%%%%%%%%%%%%%%%%%%%%%%%%%%
\section{Introduction}
\label{sec:intro}

The origin of dark matter (DM) and the cosmic matter--antimatter asymmetry remain two of the most important open problems in cosmology. At the same time, upcoming gravitational-wave (GW) observatories will probe primordial physics across frequencies that are exquisitely sensitive to small-scale curvature perturbations. Inflation provides a natural arena in which these questions can become tightly linked, because it sets both the statistics of primordial fluctuations and the initial conditions for the post-inflationary thermal history.

A particularly intriguing scenario is primordial black hole (PBH) formation from enhanced small-scale power (see Refs.~\cite{Carr:2020gox,Carr:2025kdk} for reviews). In single-field inflation, the simplest way to amplify the curvature spectrum by \(\sim 7\) orders of magnitude relative to CMB scales, as required for PBHs, is a transient non-attractor phase, typically realised as ultra-slow-roll (USR) evolution near an approximate inflection point \cite{Garcia-Bellido:2017mdw,Ezquiaga:2017fvi,Germani:2017bcs,Kannike:2017bxn,Motohashi:2017kbs,Di:2017ndc,Ballesteros:2017fsr,Pattison:2017mbe,Passaglia:2018ixg,Biagetti:2018pjj,Byrnes:2018txb,Carrilho:2019oqg,Figueroa:2020jkf,Inomata:2021tpx,Inomata:2021uqj,Pattison:2021oen,Balaji:2022rsy,Balaji:2022zur,Balaji:2022dbi,Kawai:2022emp,Karam:2022nym,Pi:2022ysn,Qin:2023lgo,Geller:2022nkr,Lorenzoni:2025gni}. During USR the inflaton velocity decays as $\dot\phi\propto a^{-3}$ and the would-be decaying mode of the curvature perturbation becomes a growing mode outside the horizon, allowing the primordial power spectrum $\mathcal{P}_\zeta$ to grow rapidly over a few e-folds \cite{Kinney:2005vj,Martin:2012pe,Byrnes:2018txb,Liu:2020oqe,Cole:2022xqc}. If the enhancement is sufficiently large, overdensities re-entering during radiation domination collapse into PBHs, which can constitute all of DM for suitable masses (see Refs.~\cite{Khlopov:2008qy,Sasaki:2018dmp,Carr:2020gox,Carr:2020xqk,Green:2020jor,Escriva:2021aeh,Villanueva-Domingo:2021spv,Escriva:2022duf,Ozsoy:2023ryl} for recent PBH reviews).

Independently, spontaneous baryogenesis offers an economical origin for the baryon asymmetry. A derivative coupling between a rolling scalar and a baryon current induces an effective chemical potential that biases equilibrium while baryon-violating reactions remain active \cite{Cohen:1987vi,Cohen:1988kt}. This immediately suggests an unavoidable interplay with USR inflation: the same dynamics invoked to produce PBH DM also controls the field velocity $\dot\phi$, and therefore the chemical potential that sources baryogenesis. Put differently, once PBHs are generated from a USR feature in a single-clock setup, it is difficult to treat the baryon asymmetry as independent.

In this work we develop a model-independent framework in which (i) the totality of DM is comprised of PBHs produced by a USR enhancement, (ii) the observed baryon asymmetry is generated by spontaneous baryogenesis sourced by the inflaton, and (iii) the same large scalar fluctuations inevitably generate an induced stochastic GW background (SGWB). The framework is predictive because PBH DM fixes the peak primordial power spectrum amplitude $\sim 10^{-2}$, while spontaneous baryogenesis depends on the inflaton velocity at the end of inflation (or reheating in the prompt-decoupling regime), since the scalar spectrum at horizon exit is related to the inflaton velocity at the same epoch. With a model for the post-peak evolution up to comoving wave numbers at the end of inflation, $k_{\rm end}$, this relation ties ${\cal P}_\zeta(k_{\rm end})$ directly to $\dot\phi_{\rm end}$, and hence to the baryon yield. Consequently, imposing both PBHs saturate the DM abundance and the inferred baryon-entropy ratio $n_B/s\simeq 8.7\times 10^{-11}$~\cite{Planck:2018jri} correlates the reheating temperature and the inflationary energy scale with the post-peak evolution of the spectrum.

The post-peak spectral slope is critical. A rapid return to the slow-roll attractor can accommodate high-scale inflation with potentially observable primordial tensor modes, whereas steeper decays generically push the scenario toward low-scale inflation to avoid baryon overproduction. This degeneracy is broken by GW astronomy: large scalar perturbations source tensor modes at second order, producing an induced SGWB that acts as a direct probe of small-scale power (see Refs.~\cite{Guzzetti:2016mkm,Caprini:2018mtu,Domenech:2021ztg} for recent reviews). For asteroid-mass PBHs the signal naturally lies in the mHz--Hz range relevant for LISA \cite{Barausse:2020rsu} and DECIGO \cite{Yuan:2021qgz,Kawamura:2020pcg}, while the high-frequency tail can enter the band of the Einstein Telescope (ET) \cite{Maggiore:2019uih}, enabling multi-band inference of the spectral shape.

We keep the analysis as model-independent as possible by adopting a phenomenological broken-power-law template for $\mathcal{P}_\zeta(k)$ that captures the universal $k^4$ rise \cite{Martin:2012pe,Liu:2020oqe,Byrnes:2018txb,Cole:2022xqc} into the USR peak and allows a variable post-peak decay. We then scan the parameter space imposing PBHs as all DM, successful spontaneous baryogenesis in the prompt-decoupling regime, and basic effective field theory (EFT) and energetic consistency requirements. We then translate the surviving region into observational targets for primordial tensors and induced GWs.

The paper is organised as follows. Sec.~\ref{sec:Inflation_Dynamics} summarises the inflationary dynamics leading to a USR phase. Sec.~\ref{sec:BPL_template} introduces the broken-power-law template for the primordial power spectrum. Sec.~\ref{sec:SB_general} presents the spontaneous baryogenesis computation and its dependence on primordial power spectrum and the reheating temperature. Sec.~\ref{sec:pbh_abundance} computes the PBH abundance and the peak amplitude required for PBHs as all DM. Sec.~\ref{sec:IGW_theory} derives the induced SGWB and assesses detectability with LISA, DECIGO and ET. Our combined results are given in Sec.~\ref{sec:results}, and finally we conclude in Sec.~\ref{sec:Discussion}.

\section{Inflationary Dynamics and the Ultra--Slow--Roll Phase}
\label{sec:Inflation_Dynamics}

To realize the formation of asteroid-mass PBHs consistent with spontaneous baryogenesis, the inflationary potential must satisfy two distinct requirements. First, on large scales $k_\textrm{CMB} \sim 0.05 \, \text{Mpc}^{-1}$, it must support a standard slow-roll phase compatible with CMB observations $P_\zeta \sim 10^{-9}$. Second, on much smaller scales, $k \sim 10^{11} - 10^{14} \, \text{Mpc}^{-1}$, the power spectrum $\mathcal{P}_\zeta$ must be enhanced by approximately seven orders of magnitude reaching $A_{\rm PBH} \sim 10^{-2}$. Such a dramatic enhancement requires a deviation from standard slow-roll dynamics. One of the simplest mechanisms to achieve this within single-field inflation is the presence of a near-inflection point in the potential, leading to a transient phase of USR inflation.

\subsection{Ultra-Slow-Roll Inflation}

The evolution of the homogeneous background field is governed by the Klein-Gordon equation
\begin{equation}
    \ddot{\phi} + 3H\dot{\phi} + V'(\phi) = 0.
    \label{eq:KG_exact}
\end{equation}
In standard slow-roll inflation, the acceleration term $\ddot{\phi}$ is negligible, and the dynamics are determined by the attractor solution $3H\dot{\phi} \simeq -V'$.
However, in the vicinity of the inflection point where $V'(\phi) \to 0$, the driving force vanishes. The friction term $3H\dot{\phi}$ must then balance the acceleration $\ddot{\phi}$, leading to the regime of USR
\begin{equation}
    \ddot{\phi} + 3H\dot{\phi} \simeq 0.
\end{equation}
Assuming the Hubble parameter $H$ is approximately constant during this brief phase, the solution for the inflaton velocity in terms of the cosmological scale factor $a$ is
\begin{equation}
    \dot{\phi}(t) \propto e^{-3H(t-t_i)} \propto a^{-3}.
    \label{eq:phidot_decay}
\end{equation}
This exponential decay of the kinetic energy defines the USR phase. It is convenient to characterize this regime using the second Hubble slow-roll parameter $\eta$ in terms of the first slow-roll parameter $\epsilon$
\begin{equation}
    \eta \equiv \frac{\dot{\epsilon}}{H\epsilon} = \frac{2\ddot{\phi}}{H\dot{\phi}} + 2\epsilon.
\end{equation}
Using Eq.~\eqref{eq:phidot_decay}, we find that $\ddot{\phi}/\dot{\phi} \simeq -3H$, implying
\begin{equation}
    \eta \simeq -6.
\end{equation}
This large, negative value of $\eta$ (in contrast to $|\eta| \ll 1$ in slow-roll) is the defining characteristic of the PBH-forming phase. It leads to the violation of the conservation of the comoving curvature perturbation $\zeta$ on super-horizon scales. As shown in Appendix.~\ref{app:USR}, during the phase where $\eta \simeq -6$, the curvature perturbation grows as $\zeta \propto a^3$. This rapid growth allows the power spectrum $\mathcal{P}_\zeta \sim |\zeta|^2$ to be amplified by a factor of $e^{6\Delta N_{\rm USR}}$, where $\Delta N_{\rm USR}$ is the duration of the USR phase.

\section{Phenomenological broken-power-law template}
\label{sec:BPL_template}

In single--field inflation with a transient USR phase, to a very good approximation the phenomenology on small scales can be captured by three scales in $k$--space and a small number of effective spectral slopes. These are (i) $k_{\rm CMB}$, the CMB pivot, where the spectrum is observed to be nearly scale invariant with amplitude $\mathcal{P}_\zeta(k_{\rm CMB}) \equiv A_s=2.1\times 10^{-9}$ and tilt $n_s=0.965$ \cite{Planck:2018jri}, (ii) $k_{\rm USR}$, the scale that exits the horizon near the end of the USR phase, where the power spectrum reaches its peak amplitude relevant for PBH formation, and (iii) $k_{\rm end}$, the scale that exits the horizon at the end of inflation, which controls the effective terminal inflaton velocity $\dot{\phi}_{\rm end}$, importantly, this sets the initial conditions for baryogenesis. 

The  broken power--law spectrum is defined with three segments, representing the pre--USR slow roll, the USR growth, and the post--peak recovery phase
\begin{equation}
\mathcal{P}_\zeta(k) \;=\;
\begin{cases}
A_s \left(\dfrac{k}{k_{\rm CMB}}\right)^{n_s-1},
& k < k_{\rm on},
\\[10pt]
A_{\rm on}\!
\left(\dfrac{k}{k_{\rm on}}\right)^{n_{\rm USR}},
& k_{\rm on} \le k < k_{\rm USR},
\\[10pt]
A_{\rm PBH}
\left(\dfrac{k}{k_{\rm USR}}\right)^{n_{\rm exit}},
& k_{\rm USR} \le k \le k_{\rm end},
\end{cases}
\label{eq:Pzeta_BPL_master}
\end{equation}
where $k_{\rm on}$ marks the onset of the growth phase, $k_{\rm USR}$ corresponds to the peak of the spectrum (end of the USR phase), and $k_{\rm end}$ marks the end of inflation. In realistic models one has $k_{\rm CMB} \ll k_{\rm USR} \ll k_{\rm end}$. We also define an effective USR growth index $n_{\rm USR}$, for a realistic transition one expects $n_{\rm USR} \simeq 4$ (see Appendix.~\ref{app:USR} for a more detailed discussion), a post--USR slow--roll tilt $n_{\rm exit}$, and the peak amplitude on PBH scales $A_{\rm PBH} \equiv \mathcal{P}_\zeta(k_{\rm USR})$.

Continuity of $\mathcal{P}_\zeta$ at $k_{\rm on}$ and $k_{\rm USR}$ fixes the intermediate amplitude $A_{\rm on}$ and scale $k_{\rm on}$ in terms of the primary parameters $(A_s,A_{\rm PBH},k_{\rm USR})$ and the slopes $(n_s,n_{\rm USR})$.

Continuity at $k_{\rm on}$ gives
\begin{equation}
A_{\rm on}
=
A_s \left(\frac{k_{\rm on}}{k_{\rm CMB}}\right)^{\!n_s-1},
\label{eq:Aon_def}
\end{equation}
while continuity at $k_{\rm USR}$ implies
\begin{equation}
A_{\rm PBH}
=
A_{\rm on}
\left(\frac{k_{\rm USR}}{k_{\rm on}}\right)^{\!n_{\rm USR}}.
\label{eq:APBH_continuity}
\end{equation}
Eliminating $A_{\rm on}$ using ~\eqref{eq:Aon_def} and \eqref{eq:APBH_continuity}, we find the transition scale $k_{\rm on}$
\begin{equation}
\frac{k_{\rm on}}{k_{\rm USR}}
=
\left[
\frac{A_{\rm PBH}}{A_s}
\left(\frac{k_{\rm CMB}}{k_{\rm USR}}\right)^{\!n_s-1}
\right]^{\!1/(n_s-1-n_{\rm USR})}.
\label{eq:kon_solution}
\end{equation}

For $k_{\rm USR} \le k \le k_{\rm end}$, the spectrum follows the post--peak recovery tail. The amplitude at the end of inflation is therefore determined by the length of this tail
\begin{equation}
\mathcal{P}_\zeta(k_{\rm end})
=
A_{\rm PBH}
\left(\frac{k_{\rm end}}{k_{\rm USR}}\right)^{\!n_{\rm exit}},
\label{eq:Pzeta_kend_from_APBH}
\end{equation}
where $n_\textrm{exit}<0$. This value is the key input for the baryogenesis computation described in the following section. %We treat $k_{\rm end}$ as a derived parameter determined by the reheating temperature.

\section{Spontaneous baryogenesis and the end of inflation}
\label{sec:SB_general}

The derivative coupling
\begin{equation}
\mathcal{L}_{\rm int}
=
\frac{1}{M_*}\,(\partial_\mu\phi)\,j_B^\mu
\label{eq:interactionlagrangian}
\end{equation}
with a baryon number current $j_B^\mu$ induces an effective baryon chemical potential $\mu_B = \dot\phi/M_*$ (see Appendix.~\ref{app:ThermalDerivation} for a detailed derivation) in a homogeneous background.
Since the coupling is suppressed by the EFT operator scale $M_*$ and we assume baryogenesis becomes relevant only at reheating, it does not affect the inflationary background evolution during the USR phase, and we can treat ${\cal P}_\zeta(k)$ and $\dot\phi$ as determined by the inflationary dynamics alone.
 While baryon number violating interactions are in equilibrium, this chemical potential biases the plasma and generates a baryon excess.

The operator in Eq.~\eqref{eq:interactionlagrangian} can be understood as the leading inflaton--SM portal in an EFT where $\phi$ has an approximate shift symmetry (broken primarily by the inflaton potential). In that case the derivative coupling is technically natural, while non-derivative couplings such as $g\,\phi\,\bar\psi\psi$ and the tower $\phi^{n+1}\bar\psi\psi/M^n$ are forbidden or spurion-suppressed by the same symmetry breaking, so they need not generate $\mathcal{O}(\phi)$ masses for SM fermions during inflation. The bias induced by $\dot\phi$ is operative only in the presence of fast charge-violating reactions. 

In this work, only reheating temperatures much larger than the electroweak scale are relevant, so electroweak sphalerons (and fast Yukawa interactions) are in equilibrium after reheating and any inflaton-induced charge asymmetry is rapidly redistributed between quark and lepton sectors. If the relevant charge-violating interactions conserve $B-L$, the late-time baryon asymmetry is controlled by the conserved charge $B-L$ once sphalerons freeze out.
In general, the decoupling temperature $T_D$ is the temperature at which baryon-violating interactions drop out of equilibrium. The frozen baryon asymmetry is (see Appendix.~\ref{app:ThermalDerivation} for a detailed derivation)
\begin{equation}
\frac{n_B}{s}
=
\frac{15 g_B}{4\pi^2 g_*}\,
\frac{\dot\phi_D}{M_* T_D},
\label{eq:nBs_basic}
\end{equation}
where $g_B \equiv \sum_i g_i B_i^2$ sums over all relativistic fermionic species carrying baryon number (with internal degrees of freedom $g_i$ and baryon charge $B_i$). For 6 Standard Model quark flavors above the electroweak scale this gives $g_B = \sum_{\rm quarks}=6\times (2\times 3)\,(1/3)^2 = 4$. This counts the total baryonic degrees of freedom while $g_*$ is the number of relativistic species at $T_D$. Here, $n_B$ denotes the net baryon-number density.% (the charge density associated with $U(1)_B$).

Our analytic treatment assumes that the baryon-number-violating reactions responsible for generating the asymmetry remain in equilibrium only briefly after reheating, so that the decoupling and reheating temperatures are approximately the same $T_D \simeq T_{\rm reh}$ and $\dot\phi_D \simeq \dot\phi_{\rm end}$. This ``prompt decoupling'' limit is physically well motivated in inflaton driven spontaneous baryogenesis because the effective chemical potential $\mu_B = \dot\phi/M_*$ redshifts rapidly once the universe becomes radiation dominated. In the absence of a sustained driving term, the homogeneous inflaton velocity scales approximately as $\dot\phi \propto a^{-3}$, implying $\mu_B/T \propto a^{-2}$. Consequently, if freeze-out occurs $\Delta N$ e-folds after reheating, the generated asymmetry is exponentially suppressed,
\begin{equation}
\left.\frac{n_B}{s}\right|_{D}\;\simeq\; e^{-2\Delta N}\left.\frac{n_B}{s}\right|_{\rm reh}\,,
\label{eq:delayed_decoupling_suppression}
\end{equation}
up to $\mathcal{O}(1)$ factors associated with the detailed post-inflationary dynamics. In the prompt-decoupling regime we focus on, the asymmetry is generated in a narrow interval around reheating and is sourced by the inflaton itself (the single clock). To leading order this ties the generated yield to a uniform-density hypersurface, strongly suppressing baryon isocurvature relative to spectator-field realizations. Observable baryon isocurvature can arise if $B$-violating reactions decouple well after reheating or if additional light fields modulate the effective rates, as discussed e.g. in Refs.~\cite{DeSimone:2016ofp,Inomata:2018htm}; we therefore treat prompt decoupling as an optimistic benchmark, noting that relaxing it reduces the yield.

The energy density stored in the induced charge densities is parametrically $\rho_{\rm asym}\sim \mu_B n_B \sim (\mu_B/T)^2 T^4$, which is negligible compared to $\rho_{\rm rad}\sim T^4$ in the relevant regime $\mu_B/T\ll 1$. Hence plasma backreaction on the inflaton background is subleading.

We remain agnostic about the UV completion of the baryon-violating sector and parametrize its departure from equilibrium by the single freeze-out scale $T_D$ defined through the interaction rate $\Gamma_{\Delta B}(T_D)\simeq H(T_D)$.
As a concrete benchmark, higher-dimensional baryon-violating operators ${\cal O}_{\Delta B}/\Lambda^{d-4}$ typically give reaction rates scaling as $\Gamma_{\Delta B}\propto T^{2d-7}/\Lambda^{2d-8}$, so that a modest change in temperature around $T\sim T_{\rm reh}$ can efficiently switch the interactions from $\Gamma_{\Delta B}\gg H$ to $\Gamma_{\Delta B}\ll H$.

The velocity of the inflaton field $\dot{\phi}$ is related to the curvature power spectrum $\mathcal{P}_\zeta(k)$ and the Hubble parameter $H$ via the relation
\begin{equation}
\mathcal{P}_\zeta(k) = \frac{H^4}{4\pi^2 \dot{\phi}^2} \,.
\label{eq:Pzeta_def}
\end{equation}
Rearranging for the velocity, we find $\dot{\phi} = H^2 / (2\pi \sqrt{\mathcal{P}_\zeta})$. We can relate the velocity at the end of inflation, $\dot{\phi}_{\rm end}$, to the velocity at CMB scales, $\dot{\phi}_{\rm CMB}$, by taking the ratio
\begin{equation}
\frac{\dot{\phi}_{\rm end}}{\dot{\phi}_{\rm CMB}}
=
\left( \frac{H_{\rm end}}{H_{\rm CMB}} \right)^2
\left[
\frac{\mathcal{P}_\zeta(k_{\rm CMB})}{\mathcal{P}_\zeta(k_{\rm end})}
\right]^{1/2} \,.
\label{eq:phidot_ratio_exact}
\end{equation}
In models of USR inflation capable of generating PBHs, the potential remains exceedingly flat between the CMB and PBH scales (typically an inflection point or plateau). Consequently, the Hubble parameter evolves slowly compared to the power spectrum, which is enhanced by some seven orders of magnitude. 

To determine the reference velocity at CMB scales $\dot{\phi}_{\rm CMB}$, we use the standard slow-roll relation $\epsilon = r/16$, where $\epsilon = \dot{\phi}^2 / (2 H^2 M_{\rm Pl}^2)$ and $r$ is the tensor-to-scalar ratio. Using $H^2 = 8\pi^2 \epsilon \mathcal{P}_\zeta M_{\rm Pl}^2$ at $k_{\rm CMB}$ we obtain

\begin{equation}
\dot{\phi}_{\rm CMB} = \frac{\pi}{4} \, r_\textrm{CMB} \, M_{\rm Pl}^2 \sqrt{\mathcal{P}_\zeta(k_{\rm CMB})} \,.
\label{eq:phidot_CMB}
\end{equation}
Substituting Eq.~\eqref{eq:phidot_CMB} into Eq.~\eqref{eq:phidot_ratio_exact}, we obtain the velocity at the end of inflation:
\begin{equation}
\dot{\phi}_{\rm end} =\frac{\pi}{4} \, r_\textrm{CMB} \, M_{\rm Pl}^2 \left(\frac{H_{\rm end}}{H_{\rm CMB}}\right)^2\frac{\mathcal{P}_\zeta(k_{\rm CMB})}{\sqrt{\mathcal{P}_\zeta(k_{\rm end})}} \,.
\label{eq:phidot_end_final}
\end{equation}
Finally, inserting this result into the baryon asymmetry expression \eqref{eq:nBs_basic} and setting $\alpha=(H_\textrm{end}/H_\textrm{CMB})^2$, yields the general result 
\begin{equation}
\frac{n_B}{s} = \frac{15 g_B }{16 \pi g_*} \left( \frac{\alpha M_{\rm Pl}^2 r_\textrm{CMB} }{M_* T_D} \right) \, \,\frac{\, \mathcal{P}_\zeta(k_{\rm CMB})}{\sqrt{\mathcal{P}_\zeta(k_{\rm end})}} \,.
\label{eq:nBs_master}
\end{equation}
For the USR case, we can substitute $P_\zeta(k_\textrm{end})$ from \eqref{eq:Pzeta_kend_from_APBH} and $P_\zeta(k_\textrm{CMB})=A_s$. The Hubble scale at the end of inflation is often approximated $H_{\rm end} \simeq H_{\rm CMB}$ , i.e., $\alpha \simeq 1$. While the Hubble rate decreases by an $\mathcal{O}(1)$ factor during the roll down the potential (typically $H_{\rm end}/H_{\rm CMB} \sim 0.5-0.7$ for plateau models)\footnote{The ratio $\alpha \equiv (H_{\rm end}/H_{\rm CMB})^2$ can be written as
\begin{equation}
\log\!\left(\frac{H_{\rm end}}{H_{\rm CMB}}\right) \;=\; -\int_{N_{\rm CMB}}^{N_{\rm end}} \epsilon(N)\, dN,
\end{equation}
using $d\log H/dN=-\epsilon$. In the USR/inflection-point scenarios relevant for PBH production, $\epsilon$ is extremely small throughout the long plateau (including the USR phase), so the integral is dominated by the final $\Delta N_{\rm end}\sim\mathcal{O}(1)$ e-fold during which $\epsilon$ grows to $\epsilon\simeq 1$ to terminate inflation. A simple estimate is therefore
\begin{equation}
\int \epsilon\, dN \;\simeq\; \bar{\epsilon}_{\rm end}\,\Delta N_{\rm end},
\end{equation}
with $\bar{\epsilon}_{\rm end}$ the average value of $\epsilon$ during the exit. Taking a conservative but typical range $\Delta N_{\rm end}\lesssim 1$ and $\bar{\epsilon}_{\rm end}\lesssim 0.7$ implies $\int \epsilon\, dN \lesssim 0.7$, i.e., $\sqrt{\alpha}=H_{\rm end}/H_{\rm CMB}\gtrsim e^{-0.7}\simeq 0.5$, i.e., $\alpha\gtrsim 0.25$. Hence $\alpha \in [0.25,1]$ captures an order-unity uncertainty in the post-plateau decrease of $H$ without assuming an unusually prolonged fast-roll end phase.}, this variation is negligible compared to the orders-of-magnitude sensitivity of the baryon asymmetry to the decoupling temperature $T_{\rm D}$ and the EFT scale $M_*$. We introduce the ratio $\alpha$ to provide model-independent benchmarks without significantly impacting the constraints. 

We note that \eqref{eq:nBs_master} captures the baryon yield generated by the inflaton velocity at the end of inflation. While the subsequent kinetic-dominated `waterfall' phase may involve larger velocities (and thus potentially enhance the yield), the dynamics in that regime are highly model-dependent. By relying on this expression, we calculate the robust, model-independent contribution determined by the power spectrum constraints, providing a guaranteed baseline for the baryon asymmetry.

\subsection{Physical Constraints and Validity of the EFT}
\label{sec:constraints}

Since we are considering instantaneous reheating, we will now refer to $T_{\textrm{reh}}$ in place of $T_D$, the effective coupling scale $M_*$, and the tensor-to-scalar ratio $r_\textrm{CMB}$ as independent parameters in our scan, physical consistency imposes a strict hierarchy of scales. To ensure the validity of the EFT description and energy conservation, the parameter space is bounded by the following conditions.

\subsubsection{Energy ceiling}
\label{sec:energyceiling}
First, the reheating temperature cannot exceed the maximum temperature corresponding to the energy density of the inflaton potential. The energy density of a relativistic plasma is $\rho_\textrm{rad} = \frac{\pi^2}{30} g_*(T_\textrm{reh}) T_\textrm{reh}^4$. By imposing energy conservation, $\rho_\textrm{rad} \leq V_\textrm{end}$. In models of USR inflation where the potential is plateau-like (as required for PBH formation), the energy density at the end of inflation is strictly less than at CMB scales. Therefore, $\rho_{\rm rad}(T_{\rm reh}) < V_{\rm end} < V_{\rm CMB}$. Hence we may use the usual slow-roll relation $V_\textrm{CMB} = \frac{3\pi^2}{2} r_\textrm{CMB} A_s M_{\rm Pl}^4$ to derive the strict upper bound
\begin{align}
    T_\textrm{reh} &< \left( \frac{45 r_\textrm{CMB} A_s}{g_*(T_\textrm{reh})} \right)^{1/4} M_{\rm Pl} \nonumber \\ & 
    \simeq 4.2 \times 10^{15} \, \text{GeV} \, \left( \frac{r_\textrm{CMB}}{0.01} \right)^{1/4} \left( \frac{106.75}{g_*(T_\textrm{reh})} \right)^{1/4} \,.
    \label{eq:TD_bound}
\end{align}
making Eq.~\eqref{eq:TD_bound} a robust ceiling. Any parameter point violating this inequality implies a reheating temperature physically unattainable because it would require more energy than exists in the system.

\subsubsection{EFT validity}
\label{sec:eftvalidity}
Second, the mass scale $M_*$ represents the cutoff of the effective operator~   \eqref{eq:interactionlagrangian}. For the EFT description to remain valid during the generation of the asymmetry, the relevant energy scales of the system, specifically the Hubble parameter during inflation, $H_{\rm inf}$, and the temperature during reheating, must lie below this cutoff. This imposes the hierarchy
\begin{equation}
    T_D\simeq T_\textrm{reh} \ll M_* \quad \text{and} \quad H_{\rm inf} < M_* \,.
    \label{eq:EFT_validity}
\end{equation}
The condition $T_\textrm{reh} < M_*$ ensures that the derivative expansion remains valid during the reheating epoch when the asymmetry freezes out. Consequently, in our numerical analysis we exclude regions where the EFT expansion breaks down, $T_{\rm reh}\gtrsim M_*$.

\subsubsection{Geometric floor}
\label{sec:geometricfloor}
 Finally, we must ensure that the inflationary era lasts long enough to encompass the USR phase responsible for generating PBHs. The comoving scale corresponding to the end of inflation, $k_{\rm end}$, is physically defined as the mode crossing the horizon at the exact moment inflation terminates, so $k_{\rm end} = a_{\rm end} H_{\rm end}$. To evaluate this scale, we invoke the instantaneous reheating approximation. We assume the universe transitions immediately from the inflationary vacuum to a radiation-dominated thermal bath. This implies continuity of the scale factor and the Hubble rate across the transition, so $a_{\rm end} \simeq a_{\rm reh}$ and $H_{\rm end} \simeq H_{\rm reh}$, allowing us to equate the horizon scale at the end of inflation with the horizon scale of the thermal bath ($k_{\rm end} \simeq k_{\rm reh}$). We calculate $k_{\rm reh} = a_{\rm reh} H_{\rm reh}$ by determining the scale factor and Hubble rate strictly from the properties of the radiation fluid. In a radiation-dominated universe, the Friedmann equation relates the Hubble rate to the energy density $\rho_{\rm rad}$. The energy density is determined by the effective degrees of freedom for energy, $g_*(T_{\rm reh})$
 
 \begin{equation}
 H_{\rm reh} = \sqrt{\frac{\rho_{\rm rad}}{3 M_{\rm Pl}^2}} = \sqrt{\frac{\pi^2}{90} g_*(T_{\rm reh})} \frac{T_{\rm reh}^2}{M_{\rm Pl}} \,.
 \label{eq:Hreh}
 \end{equation}
  We relate the scale factor at reheating to the scale factor today ($a_0 \equiv 1$) using the conservation of comoving entropy density, $s a^3 = \text{const}$. The entropy density depends on the effective degrees of freedom for entropy, $g_s(T)$
 \begin{align}
 g_s(T_{\rm reh}) T_{\rm reh}^3 a_{\rm reh}^3 &= g_{s,0} T_0^3 a_0^3  \nonumber \\ \quad \Longrightarrow \quad a_{\rm reh} &= \frac{T_0}{T_{\rm reh}} \left( \frac{g_{s,0}}{g_s(T_{\rm reh})} \right)^{1/3} \,,
  \label{eq:areh}
 \end{align}
 where $T_0 \simeq 2.35 \times 10^{-13}$ GeV is the CMB temperature today and    $g_{s,0} \simeq 3.91$ represents the current entropic degrees of freedom. Combining these expressions, the comoving horizon scale is given by $k_{\rm end} \simeq a_{\rm reh} H_{\rm reh}$, since we assume instantaneous reheating, multiplying \eqref{eq:Hreh} and \eqref{eq:areh}, it follows
 
 \begin{equation}
 k_{\rm end} \simeq \frac{T_0 T_{\rm reh}}{M_{\rm Pl}} \sqrt{\frac{\pi^2}{90}} \sqrt{g_*(T_{\rm reh})} \left( \frac{g_{s,0}}{g_s(T_{\rm reh})} \right)^{1/3} \,.
\label{eq:kend_exact}
\end{equation}
This formula is applicable to a radiation-dominated universe evolving adiabatically. For the high temperatures relevant to spontaneous baryogenesis ($T_{\rm reh} \gg 100$ GeV), the Standard Model particle content is fully relativistic, so we may set $g_*(T_{\rm reh}) \approx g_s(T_{\rm reh}) \simeq 106.75$. Evaluating the physical constants yields the precise linear mapping

\begin{equation}k_{\rm end} \simeq 1.7 \times 10^{14} \, \text{Mpc}^{-1} \left( \frac{T_{\rm reh}}{10^7 \, \text{GeV}} \right) \left( \frac{g_*(T_{\rm reh})}{106.75} \right)^{1/6} \,.\label{eq:kend_numerical}\end{equation}
This instantaneous reheating mapping can be substituted into Eq.~\eqref{eq:nBs_master} to obtain $n_B/s$ in terms of the reheating temperature. We also find, that it imposes a `geometric floor' on the parameter space. For the model to be self-consistent, the inflationary phase must sustain modes up to $k_{\rm end}$, which necessitates that the PBH-generating scale $k_{\rm USR}$ exited the horizon \textit{before} inflation ended. This requires
\begin{equation}
k_{\rm end}(T_{\rm reh}) > k_{\rm USR} \,.
\label{eq:kend_limit}\end{equation}
If $T_{\rm reh}$ is too low, $k_{\rm end}$ drops below $k_{\rm USR}$. This physically implies that inflation ended before the field could reach the spectral feature responsible for PBH formation, rendering the scenario invalid. A non-standard post-inflationary equation of state (e.g. an early matter-dominated or kination phase) would rescale the relation between comoving scales and temperatures and can modify the resulting phenomenology.

\section{PBH Formation and Dark Matter Abundance}
\label{sec:pbh_abundance}

To constrain the model parameters, we require that PBHs constitute the entirety of the observed DM energy density ($f_{\text{PBH}} = 1$). The calculation proceeds in four steps: relating the comoving wavenumber $k$ to the PBH mass $M_\textrm{PBH}$, computing the variance of density perturbations $\sigma^2(k)$, determining the collapse fraction $\beta(k)$, and integrating the resulting mass function.

First, we relate the PBH mass to the comoving wavenumber $k$ at horizon re-entry. Assuming standard radiation domination, the PBH mass is given by~\cite{Ozsoy:2023ryl}
\begin{align}
M_{\rm PBH}(k) &= 30M_\odot 
\left(\frac{3.2\times 10^{5}\,\mathrm{Mpc}^{-1}}{k}\right)^{2} \\ \nonumber
& \qquad \qquad \qquad \quad \times \left(\frac{\gamma_c}{0.2}\right) 
\left(\frac{g_*(T_f)}{106.75}\right)^{-1/6},
\label{eq:Mpbh}
\end{align}
where $g_*(T_f)$ is the number of effectively massless degrees of freedom at the time of PBH formation and $\gamma_c = 0.2$ is the typical formation efficiency factor.

We evaluate the variance of the smoothed density field at the horizon scale $R = k^{-1}$
\begin{equation}
\sigma^2(k) = \frac{16}{81} \int d\log k' \, \mathcal{P}_\zeta(k') 
\left(\frac{k'}{k}\right)^4 W^2(k'/k) \, T^2(k'/k).
\label{eq:sigma2}
\end{equation}
Here we apply a Gaussian window function $W(x) = e^{-x^2/2}$ and the radiation-era transfer function
\begin{equation}
T(x) = \frac{9\sqrt{3}}{x^3}\left[\sin\left(\frac{x}{\sqrt{3}}\right) 
- \frac{x}{\sqrt{3}}\cos\left(\frac{x}{\sqrt{3}}\right)\right],
\label{eq:transfer}
\end{equation}
following Ref.~\cite{Kawasaki:2019mbl}.

We estimate the PBH abundance using the Press-Schechter formalism. The present-day fraction of DM in PBHs is
\begin{align}
f_{\rm PBH} &= \int d\log k \, 
\frac{\sqrt{M_{\rm eq}}}{\Omega_{\rm CDM}^{\rm eq}} 
\left(\frac{1}{M_{\rm PBH}(k)}\right)^{1/2} \nonumber\\
&\quad\quad\quad\quad\quad\quad\quad\quad\quad\quad\times \mathrm{erfc}\!\left(\frac{\delta_c}{\sqrt{2\sigma^2(k)}}\right),
\label{eq:fpbh}
\end{align}
where the factor of $1/2$ in the standard collapse fraction $\beta = \frac{1}{2}\,\mathrm{erfc}(\delta_c/\sqrt{2}\sigma)$ cancels against the Jacobian $|d\log M/d\log k| = 2$ when changing integration variables. In practice a localized peak in ${\cal P}_\zeta(k)$ produces a narrow extended PBH mass function, and the integral over $\log k$ in Eq.~\eqref{eq:fpbh} accounts for its finite width.

We adopt Planck 2018 cosmological parameters~\cite{Planck:2018vyg} for current DM energy density fraction $\Omega_{\rm CDM,0} = 0.265$ and the redshift at matter-radiation equality $z_{\rm eq} = 3402$, implying $\Omega_{\rm CDM}^{\rm eq} = 0.42$. The horizon mass at matter-radiation equality is $M_{\rm eq} = 2.94 \times 10^{17}\,M_\odot$, and we use the critical density threshold $\delta_c = 0.45$~\cite{Carr:1975qj,Musco:2012au,Harada:2013epa}. The Press-Schechter formalism involves $\mathcal{O}(1)$ uncertainties  from the choice of window function, threshold value, and corrections for non-linear effects. However, these uncertainties are subdominant compared to the exponential sensitivity of $\beta$ to $\delta_c/\sigma$. We emphasize that our PBH abundance results use a Gaussian benchmark for the statistics of $\zeta$. In USR realizations the non-attractor phase can generate significant non-Gaussianity, which can shift the required peak amplitude $A_{\rm PBH}$ for a fixed $f_{\rm PBH}$; we consider Gaussian statistics as an indicative baseline, and refer the reader to e.g. Refs.~\cite{Gow:2020bzo,Gow:2022jfb} for careful treatments of these effects. Additionally, non-linear shape and threshold effects can be important to consider when doing precision amplitude calculations, we refer to e.g. Refs.~\cite{Musco:2018rwt,Ferrante:2022mui} for more detailed discussion of these effects.

By enforcing $f_{\rm PBH} = 1$, we determine the required peak amplitude $A_{\rm PBH}$ for PBH masses in the asteroid-mass window ($10^{-16}\,M_\odot \lesssim M \lesssim 10^{-10}\,M_\odot$), corresponding to $k_{\rm USR} \in [10^{11}, 10^{14}]\,\mathrm{Mpc}^{-1}$~\cite{Carr:2020gox,Carr:2020xqk,Green:2020jor,DelaTorreLuque:2024qms}.
\section{Induced Gravitational Waves}
\label{sec:IGW_theory}

In addition to the production of PBHs, the large curvature perturbations required to collapse overdensities inevitably source a SGWB at second order in cosmological perturbation theory. Unlike primary GWs from inflation (which scale linearly with the tensor-to-scalar ratio $r$), these induced waves are sourced by quadratic terms of the curvature perturbation $\mathcal{P}_\zeta$.

\subsection{Current Energy Density}
The induced GWs are described by their energy density per logarithmic frequency interval, normalized by the critical density. Assuming the waves are generated during the radiation-dominated epoch at a temperature $T_c$, the abundance today is determined by the redshift of radiation and the change in effective degrees of freedom~\cite{Domenech:2021ztg}
\begin{equation}
    \Omega_{\rm GW,0} h^2 = \Omega_{r,0}h^2 \left( \frac{g_* (T_c)}{g_{*,0}} \right) \left( \frac{g_{*s} (T_c)}{g_{*s,0}} \right)^{-4/3} \Omega_{\rm GW,c} \,.
\end{equation}
Here, $\Omega_{r,0}h^2 \approx 4.18 \times 10^{-5}$ is the present-day radiation density~\cite{Planck:2018vyg} and $\Omega_{\rm GW,c}$ is GW spectral density at the time of formation. The quantities $g_*$ and $g_{*s}$ denote the effective degrees of freedom for energy density and entropy, respectively. Today, these values are $g_{*,0} \approx 3.36$ and $g_{*s,0} \approx 3.91$. 

For the PBH mass range of interest (asteroid-mass PBHs), the relevant modes re-enter the horizon deep in the radiation era when the Standard Model degrees of freedom are fully excited ($g_* \simeq g_{*s} \simeq 106.75$). In this regime, the relation simplifies to 
\begin{equation}
    \Omega_{\rm GW,0} h^2 \simeq 1.62 \times 10^{-5} \, \Omega_{\rm GW,c} \,.
        \label{eq:Omega_GW_today}
\end{equation}

\subsection{The Tensor Source at Formation}
The spectral density at the time of formation, $\Omega_{\rm GW,c}$, is defined in terms of the time-averaged tensor power spectrum $\overline{\mathcal{P}_h}$ as
\begin{equation}
    \Omega_{\rm GW,c}(k, \tau) = \frac{1}{24} \left( \frac{k}{aH} \right)^2 \overline{\mathcal{P}_h(k,\tau)} \,,
    \label{eq:Omega_c_def}
\end{equation}
where the overline denotes oscillation averaging and $\tau$ denotes conformal time. The tensor perturbations $h_\lambda$ (for polarizations $\lambda = +, \times$) obey the equation of motion sourced by scalar perturbations
\begin{equation}
    h''_\lambda(\mathbf{k},\tau) + 2\mathcal{H}h'_\lambda(\mathbf{k},\tau) + k^2 h_\lambda(\mathbf{k},\tau) = 4 S_\lambda(\mathbf{k},\tau) \,.
\end{equation}
where $\mathcal{H} = aH = 2/ [(1+3w)\tau]$ and $w$ determines the equation-of-state of the fluid that fills the universe. This equation is solved using the Green's function method~\cite{Ananda:2006af,Baumann:2007zm}. Then the two-point correlation function of the source term can be calculated allowing us to express the power spectrum $\mathcal{P}_h$ as a convolution of the scalar power spectrum $\mathcal{P}_\zeta$. Evaluating this source term requires integrating over the momentum modes of the scalar perturbations. For numerical evaluation, we adopt the dimensionless $s$ and $t$ variables introduced in Ref.~\cite{Kohri:2018awv}, the tensor spectrum from Ref.~\cite{Kohri:2018awv} then reads
\begin{align}
\label{eq:powerspectrum}
P_h(k, \tau) =& 2\int_{0}^{\infty} dt \int_{-1}^{1} ds \left[ \frac{t (2 + t) (s^2 - 1)}{(1 - s + t) (1 + s + t)}\right]^2 \\\nonumber
&\quad \times {\cal P}_{\zeta}\left(\frac{k (t+s+1)}{2}\right) {\cal P}_{\zeta}\left(\frac{k (t-s+1)}{2}\right) \\\nonumber
&\quad \times I_{\rm RD}^2(s,t, k\tau).
\end{align}

In the late-time limit of the radiation dominated universe, i.e., where $k\tau\rightarrow\infty$, we use the oscillation averaged result
\begin{align}
\label{eq:GWtf}
&\overline{I^2_{\textrm{RD}}(s,t, k\tau \to \infty)} =  \frac{288 (-5 + s^2 + t (2 + t))^2}{(1 - s + t)^6 (1 + s + t)^6} \times
\\
&\qquad\Bigg\{ \frac{\pi^2}{4} \left(-5 + s^2 + t (2 + t)\right)^2 \Theta\left(t-(\sqrt{3}-1)\right) +
\nonumber \\\nonumber
&\qquad \Bigg[-(t - s + 1) (t + s + 1) + \\ \nonumber & \qquad 
\frac{1}{2} (-5 + s^2 + t (2 + t))  \log\left|\frac{(-2 + t (2 + t))}{3 - s^2}\right|\Bigg]^2\Bigg\},
\end{align}
where $\Theta$ is the usual Heaviside theta function. Hence the averaged analytical transfer function during radiation domination \eqref{eq:GWtf} above can be substituted into \eqref{eq:powerspectrum}. The resulting integral will yield the oscillation averaged power spectrum $\overline{P_h(k,\tau)}$ which can be substituted into Eq.~\eqref{eq:Omega_c_def}. Finally \eqref{eq:Omega_c_def} can be substituted into ~\eqref{eq:Omega_GW_today}, and the resulting present day dimensionless GW background density can be compared with experimental projections.

\subsection{Detectability and Signal-to-Noise Ratio}
To determine the observability of these signals, we calculate the Signal-to-Noise Ratio (SNR) for future interferometers such as LISA~\cite{Schmitz:2020syl}, DECIGO~\cite{Yagi:2011wg}, and the ET~\cite{Maggiore:2019uih}. The optimal broadband SNR for a stochastic background, integrated over frequency for an observation time $T_{\rm obs}$, is given by~\cite{Thrane:2013oya}
\begin{equation}
    \text{SNR} = \sqrt{ 2 T_{\rm obs} \int_{f_{\min}}^{f_{\max}} df \, \left[ \frac{\Omega_{\rm GW,0}(f) h^2}{\Omega_{\rm noise}(f) h^2} \right]^2 } \,,
    \label{eq:SNR_def}
\end{equation}
where $[f_{\min}, f_{\max}]$ defines the detector's bandwidth. The effective noise energy density $\Omega_{\rm noise}(f) h^2$ is derived from the detector's strain sensitivity $S_n(f)$ via:
\begin{equation}
    \Omega_{\rm noise}(f) h^2 = \frac{2\pi^2}{3 H_0^2} f^3 S_n(f) h^2 \,.
        \label{eq:noise}
\end{equation}
After the effective noise energy density is obtained from each detector's strain sensitivity curve, it can be substituted into \eqref{eq:SNR_def} along with the present day dimensionless GW background density in Eq.~\eqref{eq:Omega_GW_today} to obtain the SNR. Naively, detection of $\text{SNR} \gg 1$ are expected to be resolvable. We will assume the baseline instrumental sensitivity curves and do not include additional astrophysical foreground subtraction uncertainties. In Sec. \ref{subsec:GW_results}, we numerically evaluate the SNRs for the specific spectral shapes required for baryogenesis and PBH DM. For a given mode in Fourier space, the frequency of GWs today is given by the useful relation $f\simeq1.55\times 10^{-15} \left(k/\textrm{Mpc}^{-1}\right) \, \textrm{Hz}$.

\section{Results}
\label{sec:results}
\subsection{Power Spectrum Amplitude Required for PBH DM}
First, we compute the power spectrum amplitude required to enable PBH DM which we can then use to compute the allowable parameter region that simultaneously produces the observed baryon asymmetry of the universe. This imposes tight constraints on the energy scale of inflation. We begin by determining the necessary enhancement of the power spectrum. The required peak amplitude $A_{\text{PBH}}$ depends on the steepness of the post-peak decay, parameterized by the spectral index $n_{\text{exit}}$. Steeper tails (e.g., $n_{\text{exit}} \lesssim -2$) confine the variance $\sigma^2$ more strictly to the peak region, necessitating a higher amplitude compared to the standard slow-roll attractor case. 

\begin{figure}[h]
    \centering
    \includegraphics[width=0.45\textwidth]{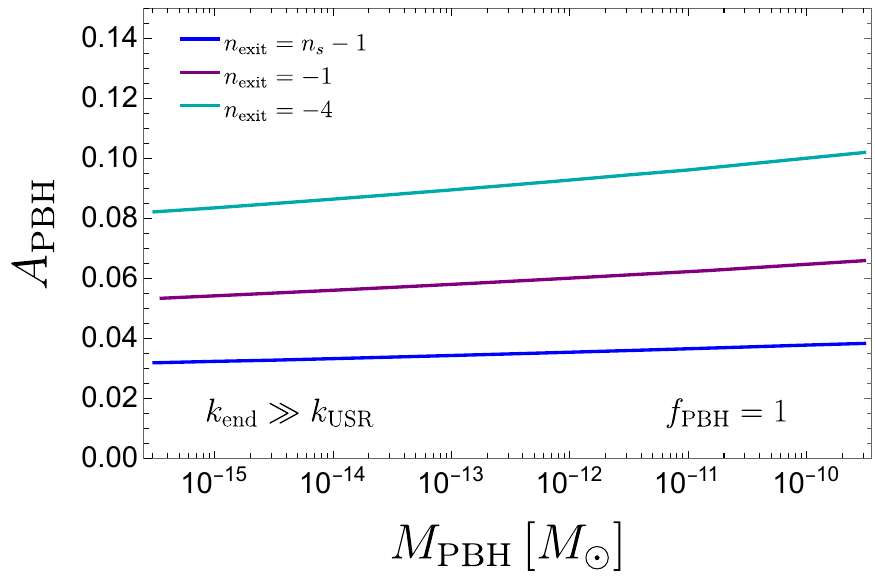}
    \caption{The peak amplitude $A_\textrm{PBH}$ as a function of PBH mass required for PBHs to saturate the DM density, i.e., $f_\textrm{PBH}=1$ assuming a narrow mass function for various inflationary spectral decay indices after the ultra-slow-roll phase. }
    \label{fig:APBH}
\end{figure}

By setting Eq.~\eqref{eq:fpbh} to unity to saturate the DM abundance with PBHs and solving for $A_\textrm{PBH}$, for various spectral indices and $k_\textrm{USR}$ values, we get Fig.~\ref{fig:APBH}. We assume that $k_\textrm{end}\gg k_\textrm{USR}$ to ensure that the end of inflation occurs sufficiently far from the USR peak. If the end of inflation is very close to the end of the USR phase, this would truncate the number of UV $k$-modes contributing to $\sigma^2$, meaning a larger amplitude would be required to saturate the DM abundance for the same $k_\textrm{USR}$. However, these changes are $\mathcal{O}(1)$ at most and therefore do not significantly affect the parameter regions for scales relevant for baryogenesis, which span orders of magnitude as will be shown in the next section.

In this figure, we see that for $n_{\text{exit}} \approx n_s-1\simeq-0.04$, meaning the inflaton settles back to the slow-roll attractor after the USR phase, this requires $A_\textrm{PBH}\simeq 0.030 - 0.038$. For $n_{\text{exit}} = -1$ (intermediate decay) $A_\textrm{PBH}\simeq 0.053 - 0.066$  and finally for $n_{\text{exit}} \le -4$ (steepest decay) $A_\textrm{PBH} \simeq 0.082 - 0.101$.  The marginal increase in the $A_\textrm{PBH}$ required for higher masses reflects the fact that lighter PBHs contribute more to the total DM fraction since $\Omega_\textrm{PBH}\propto \frac{\beta}{M_\textrm{PBH}^{1/2}}$ where $\beta$ is the fraction of collapsed horizons at formation.

\subsection{Spontaneous Baryogenesis Parameter Space}
\begin{figure*}[t]
    \centering
    % Top Row: Flat Spectrum
    \begin{minipage}[b]{0.48\textwidth}
        \centering
        \includegraphics[width=\textwidth]{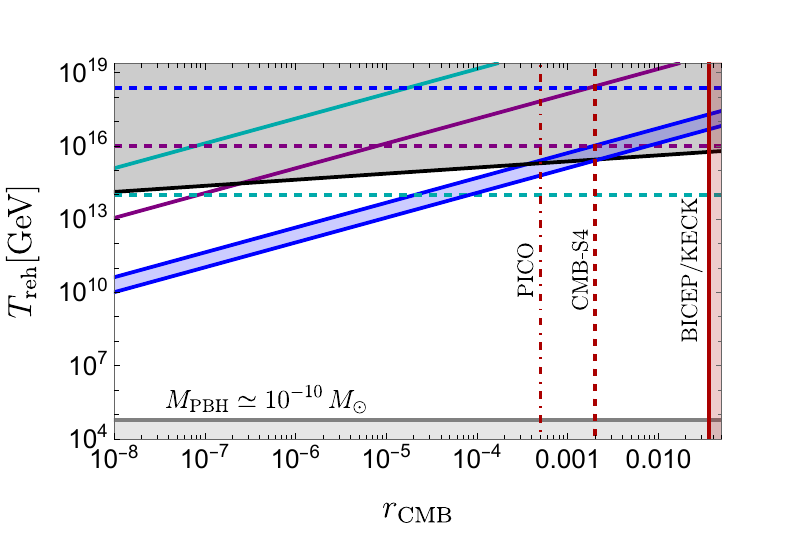}
        \par\vspace{2mm} % Add a little space between image and caption
        (a) Flat-tail, $k_{\rm USR} = 10^{11} \, \text{Mpc}^{-1}$
        \label{fig:flat_lowk}
    \end{minipage}
    \hfill
    \begin{minipage}[b]{0.48\textwidth}
        \centering
        \includegraphics[width=\textwidth]{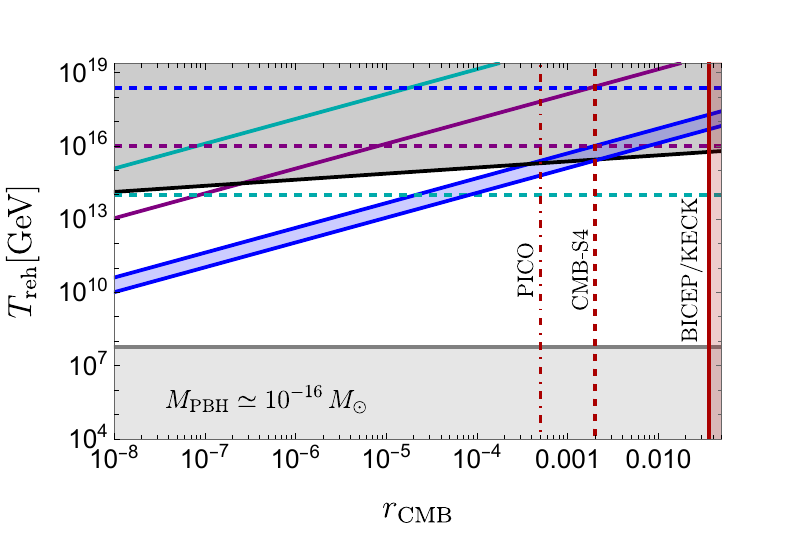}
        \par\vspace{2mm}
        (b) Flat-tail, $k_{\rm USR} = 10^{14} \, \text{Mpc}^{-1}$
        \label{fig:flat_highk}
    \end{minipage}
    
    \vspace{0.5cm} % Vertical separation between rows
    
    % Bottom Row: Steep Spectrum
    \begin{minipage}[b]{0.48\textwidth}
        \centering
        \includegraphics[width=\textwidth]{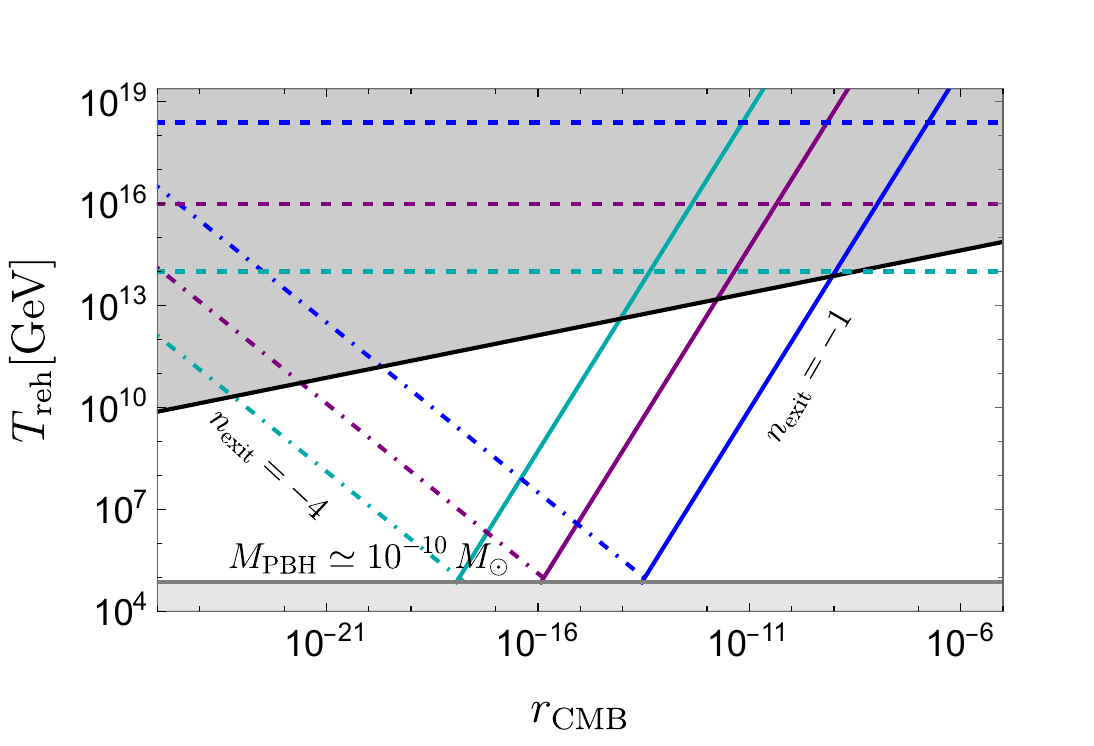}
        \par\vspace{2mm}
        (c) Steep-tail, $k_{\rm USR} = 10^{11} \, \text{Mpc}^{-1}$
        \label{fig:steep_lowk}
    \end{minipage}
    \hfill
    \begin{minipage}[b]{0.48\textwidth}
        \centering
        \includegraphics[width=\textwidth]{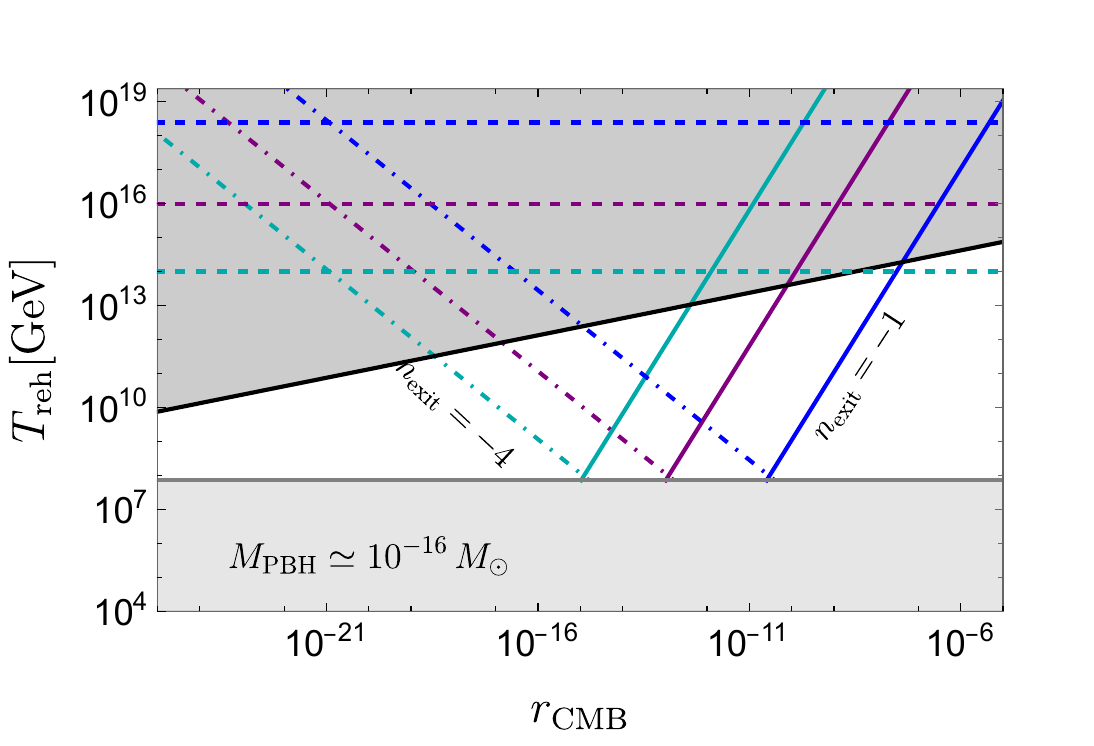}
        \par\vspace{2mm}
        (d) Steep-tail, $k_{\rm USR} = 10^{14} \, \text{Mpc}^{-1}$
        \label{fig:steep_highk}
    \end{minipage}
    
    \caption{The parameter space for spontaneous baryogenesis compatible with $100\%$ PBH DM. The contours indicate the required reheating temperature $T_{\rm reh}$ versus the tensor-to-scalar ratio $r_\textrm{CMB}$ for different effective coupling scales $M_*$ of $M_{\rm Pl}$ (blue), $10^{16}$ GeV (purple) and $10^{14}$ GeV (cyan). The lower part of the blue region corresponds to $\alpha=0.25$ while the upper part corresponds to $\alpha=1$. The grey shaded regions represent physical exclusion zones: the upper region violates energy conservation ($T_{\rm reh} > V^{1/4}$), and the lower region violates the geometric duration of inflation ($k_{\rm end} < k_{\rm USR}$). Vertical lines indicate the current exclusion limit from BICEP/Keck ($r_\textrm{CMB} < 0.036$) and the projected sensitivity of future experiments like CMB-S4 ($r_\textrm{CMB} < 2\times 10^{-3}$) and PICO ($r_{\rm CMB}< 5\times 10^{-4}$). \textbf{Top Row:} Results for the standard slow-roll attractor tail ($n_{\rm exit} \approx -0.04$). \textbf{Bottom Row:} Results for steeper post-peak decays ($n_{\rm exit} = -1, -4$), which force the solution into a low-scale inflation regime with extremely suppressed $r_{\rm CMB}$.}
    \label{fig:results_scan}
\end{figure*}
Having computed the requirements for PBH DM in the $(M_\textrm{PBH},A_\textrm{PBH})$ plane in Fig.~\ref{fig:APBH}, we can now determine the parameter region for successful spontaneous baryogenesis. As mentioned in previous sections, throughout this analysis, we assume instantaneous reheating and identify the baryon freeze-out temperature $T_D$ with the reheating temperature $T_{\rm reh}$ ($T_D \simeq T_{\rm reh}$). By solving \eqref{eq:nBs_master} set to the observed baryon-to-entropy ratio $n_B/s\simeq8.7\times 10^{-11}$ \cite{Planck:2018jri} for $T_\textrm{reh}$ as a function of scale $M_*$ and $r_\textrm{CMB}$, we obtain the viable parameter space in Fig.~\ref{fig:results_scan}. The plots in Fig.~\ref{fig:results_scan} display the required reheating temperature $T_{\rm reh}$ as a function of $r_\textrm{CMB}$ to satisfy the observed baryon asymmetry. We consider three representative values for the effective coupling scale $M_*$: the Planck scale ($M_{\rm Pl}$, blue), a GUT-scale suppression ($10^{16}$ GeV, purple), and an intermediate scale ($10^{14}$ GeV, cyan).

The results of our numerical scan reveal a tightly constrained ``wedge" of viability. The favoured parameter space in Fig.~\ref{fig:results_scan} represent parameter combinations where the baryon asymmetry is generated at the observed level without violating physical consistency conditions. The viable region is bounded by two hard constraints. From above, the energy ceiling constraint in Sec.~\ref{sec:energyceiling} ensures that the reheating temperature cannot exceed the energy scale of inflation. High reheating temperatures are only permitted if the inflationary energy scale (and thus $r_\textrm{CMB}$) is sufficiently large, this is shown in dark gray. From below, there is a geometric floor (shown in light gray) as described in Sec.~\ref{sec:geometricfloor}, this ensures inflation lasts long enough to include the USR peak. If $T_{\rm reh}$ is too low, the total number of e-folds $N_{\rm tot}$ decreases, causing the end-of-inflation scale $k_{\rm end}$ to unphysically retreat below the PBH scale $k_{\rm USR}$. This floor rises significantly for higher peak wavenumbers (comparing the left panels to the right panels). This is expected as a late-time peak at $k_{\rm USR}=10^{14}\,\text{Mpc}^{-1}$ requires a more prolonged inflationary epoch (and thus a higher minimum $T_{\rm reh}$) than a peak at $k_\textrm{USR}=10^{11}\,\text{Mpc}^{-1}$. The parameter region for baryogenesis enforces an effective chemical potential that satisfies $\mu_B/T \ll 1$, ensuring the validity of the linearized expansion shown in Appendix.~\ref{app:ThermalDerivation}.

In the top panels of Fig.~\ref{fig:results_scan}, the power spectrum returns to the standard slow-roll attractor ($n_{\rm exit} \approx n_s - 1 \approx -0.04$) immediately after the PBH peak. The left and right panels correspond to a PBH DM mass of $\simeq 10^{-10}M_\odot$ and $\simeq 10^{-16}M_\odot$ respectively. The solid coloured lines show the required $T_\textrm{reh}$ and $r_\textrm{CMB}$ to get the correct baryon asymmetry for different EFT scales for the baryogenesis operator $M_*$. Points below (above) each curve correspond to a larger (smaller) baryon yield than observed. The colored dashed lines correspond to the EFT validity constraint $T_\textrm{reh}<M_*$ outlined in Sec.~\ref{sec:eftvalidity}, only the regions where the solid lines lie below the intersection of the same colored dashed line are allowed. For $M_*=M_\textrm{Pl}$ we show a shaded region, where the upper part corresponds to $\alpha=1$ and the lower part corresponds to $\alpha=0.25$, corresponding to the typical ranges favored by USR models of $0.25\lesssim (H_\textrm{end}/H_\textrm{CMB})^2 \lesssim 1$. Hence for the flat-tail case, we see that the allowable region favors $r_\textrm{CMB} \lesssim 2\times 10^{-3}$ and $T_\textrm{reh}\lesssim 10^{15}$ GeV and larger $M_*$. Smaller values of $M_*$ are in tension with the energy ceiling, requiring larger $T_\textrm{reh}$ or smaller $r_\textrm{CMB}$ to obtain the same baryon asymmetry as shown in cyan and purple.

The bottom panels of Fig.~\ref{fig:results_scan} illustrate the dramatic impact of steeper post-peak decays ($n_{\rm exit} = -1$ and $-4$). These benchmarks bracket the physically realized behavior of the inflaton during the exit from the USR phase: the intermediate decay ($n_{\rm exit} = -1$) represents gradual transitions typical of inflection-point models, while the steepest decay ($n_{\rm exit} = -4$) corresponds to an extreme baseline for single-field inflation.

Steep spectral tails enhance the baryon yield by retaining an imprint of the large USR velocities. To prevent overproduction of baryons, the overall normalization of the power spectrum must be suppressed. Since the scalar amplitude $A_{\rm PBH}$ is fixed by the DM abundance, the only remaining parameter for regulation is the tensor-to-scalar ratio $r_\textrm{CMB}$. The results confirm this scaling behavior in the intermediate decay ($n_{\rm exit} = -1$) case, the contours shift to lower $r_\textrm{CMB}$ values ($\lesssim 10^{-9}$) and reheating temperatures $T_{\rm reh} \lesssim 10^{13}$ GeV within a low-scale inflation framework. For the steepest decay scenario, the enhancement is so severe that $r_\textrm{CMB}$ must be suppressed to extremely small values ($< 10^{-12}$). Consequently, the allowed inflationary energy scale drops to $V^{1/4} \lesssim 10^{12}$ GeV. In this regime, the geometric floor becomes the dominant constraint.\footnote{Strictly speaking, as the reheating temperature drops towards the geometric floor ($k_{\rm end} \to k_{\rm USR}$), the integrated variance decreases due to the cutoff, requiring a slight increase in $A_{\rm PBH}$ to maintain $f_{\rm PBH}=1$. We neglect this secondary correction as it affects only a negligible strip of parameter space immediately adjacent to the geometric floor exclusion zone.} For $k_{\rm USR}=10^{14}\,\text{Mpc}^{-1}$, the $n_{\rm exit}=-4$ solution is the most strongly constrained, because the low $T_{\rm reh}$ required to obtain the baryon asymmetry is in tension with the large values required to sustain the required duration of inflation.

The vertical lines in the top panels demonstrate that the flat-tail scenario with large EFT scale $M_*$ is most testable via tensor-to-scalar ratio measurements at CMB scales. For a Planck-suppressed $M_* = M_{\rm Pl}$ the allowable region is potentially within the reach of next-generation CMB experiments. The red solid vertical line shows the BICEP/KECK constraint of $r_\textrm{CMB}<0.036$ \cite{BICEP:2021xfz}, the red-dashed line shows the prospective CMB-S4 sensitivity of $r_\textrm{CMB}<0.002$ \cite{CMB-S4:2016ple} and finally the red dot-dashed shows the projected PICO sensitivity of $r_\textrm{CMB}<5\times 10^{-4}$ \cite{NASAPICO:2019thw}. We see that with CMB-S4 marginally sensitive to the flat-tail case with $\alpha\simeq 0.25$ while PICO is marginally sensitive to this scenario even at $\alpha\rightarrow1$ for  $T_\textrm{reh}\sim 10^{14}$ GeV. 

In summary, the detectability of the tensor-scalar ratio signal acts as a discriminator for the inflationary dynamics. While the standard ``flat" tail scenario predicts  high-scale inflation marginally detectable by future CMB missions at $T_\textrm{reh}\simeq10^{14}$ GeV, any mechanism that sharpens the spectral tail ($n_{\rm exit} \lesssim -1$) effectively enforces a low-scale inflation scenario ($r_\textrm{CMB} \ll 10^{-9}$), invisible to CMB probes but potentially accessible via the induced GW background which will be discussed in Sec.~\ref{subsec:GW_results}.

\subsection{Gravitational Wave Detectability}
\label{subsec:GW_results}

The primordial GW spectrum is mostly unconstrained on the much shorter length-scale scalar perturbations responsible for PBH formation. We calculate the expected contribution to the SGWB from PBH formation in our models, and find that the signal overlaps significantly with the projected sensitivity for several next-generation detectors like LISA \cite{Barausse:2020rsu}, the ET \cite{Maggiore:2019uih}, and DECIGO \cite{Yuan:2021qgz,Kawamura:2020pcg}. The large amplitude scalar perturbations required for PBH formation inevitably source an induced SGWB. Using the formalism outlined in Sec.~\ref{sec:IGW_theory}, we calculate the SNR for the experiments using \eqref{eq:SNR_def} assuming 4 years of observation time for LISA and 1 year for ET and DECIGO.

\begin{figure}[h!]
    \centering
    \includegraphics[width=0.5\textwidth]{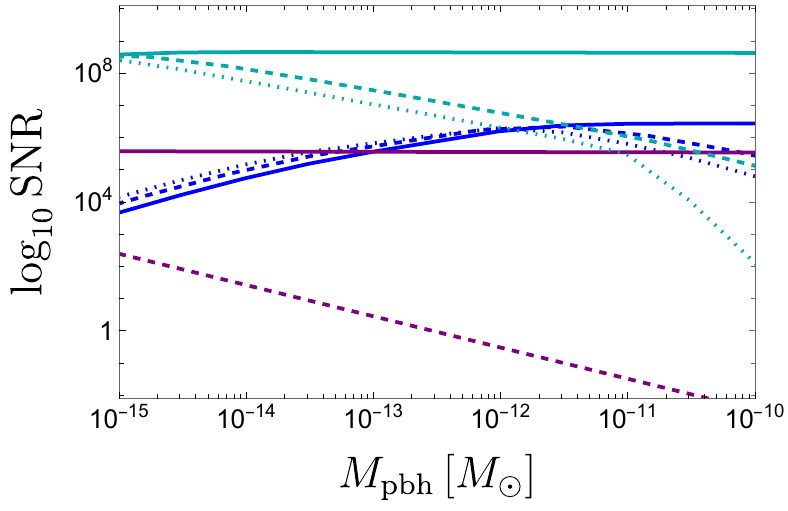}
    \caption{The predicted SNRs for LISA (blue), ET (purple), and DECIGO (cyan) as a function of the PBH mass $M_{\rm PBH}$ for 100\% PBH DM. The different line styles correspond to different post-peak spectral slopes: flat-tail ($n_{\rm exit} =n_s-1$, solid), intermediate decay ($n_{\rm exit} = -1$, dashed), and steep decay ($n_{\rm exit} = -4$, dotted). }
    \label{fig:GW_SNR}
\end{figure}

In Fig.~\ref{fig:GW_SNR}, we plot the predicted SNR as a function of the PBH mass $M_{\rm PBH}$ across the asteroid-mass window where 100\% PBH DM is possible. This corresponds to the $(M_\textrm{PBH},A_\textrm{PBH})$ values shown in Fig.~\ref{fig:APBH}. We focus on sufficiently large $T_{\rm reh}$ such that $k_{\rm end}(T_{\rm reh})\gg k_{\rm USR}$, so that the UV cutoff of the spectrum lies far beyond the frequency bands of these detectors. This is consistent with the viable baryogenesis parameter space identified in Fig.~\ref{fig:results_scan}.

Across the entire PBH DM mass range for the flat and intermediate decay case, the predicted signal is exceptionally loud. LISA achieves $10^3\lesssim\text{SNR} \lesssim 10^7$, while DECIGO reaches $10^4 \lesssim\text{SNR} \lesssim 10^8$. LISA shows minimal dependence on the decay index because the strength of the signal is dominated by the infrared tail and peak of the GW spectrum. An interesting observation is that larger SNRs for steeper decay indices are found for lower PBH masses $\lesssim 10^{-12}M_\odot$, this is simply because the required amplitude $A_\textrm{PBH}$ for the the steeper tail cases is slightly increased, as shown in Fig.~\ref{fig:APBH}, since the GW energy density has $\propto A_\textrm{PBH}^2$ dependence. In cases where the peak of the spectrum dominates the SNR, as is the case for LISA, this can create a marginally louder signal. For DECIGO, with maximal sensitivity at higher frequencies, there is a stronger dependence on the decay index, with larger GW suppression at higher PBH masses (lower associated $k_\textrm{USR})$. For the ground-based ET, in the flat-tail case, the signal is broadband and scale-invariant. Even when the peak is at low frequencies (high $M_{\rm PBH}$), ET detects the high-frequency tail with $\text{SNR} \sim 10^5$. For the intermediate tail, the signal is marginally observable at higher frequencies but completely invisible for the steep decay case. This is because, in the steep decay case, the UV tail of the GW spectrum is severely suppressed in the ET sensitivity band when requiring 100\% PBH DM.
   
This demonstrates ET's application as a spectral discriminator, it detects the broadband signal of the flat-tail but loses sensitivity to the sharp cutoff of the steep tail for high PBH masses. Consequently, a multi-band observation can distinguish the inflationary dynamics: a detection in LISA or DECIGO combined with a non-detection in ET would strongly favor a steep spectral decay ($n_{\rm exit} \lesssim -4$), implying a sharp feature in the inflationary potential, consistent with the low-scale inflation scenarios required by our baryogenesis analysis.

\section{Discussion and Conclusions}
\label{sec:Discussion}
In this work, we have proposed a unified framework in which a single period of USR inflation simultaneously accounts for the total DM abundance via PBHs, the baryon asymmetry of the universe via spontaneous baryogenesis, and a SGWB detectable by future interferometers. The central result of our analysis is that these three distinct phenomena are intrinsic consequences of the same inflationary dynamics. The exponential suppression of the inflaton velocity, required to enhance small-scale curvature perturbations for PBH formation, unavoidably sets the boundary conditions for the derivative coupling driving baryogenesis. Consequently, once the peak amplitude is fixed to satisfy the DM constraint, the resulting baryon asymmetry ceases to be a free parameter and becomes a predictive function of the inflationary energy scale, post-USR decay index and reheating temperature.

Our exploration of the parameter space reveals that the viability of this mechanism depends critically on the post-peak evolution of the power spectrum. If the spectrum returns to the standard slow-roll attractor immediately after the PBH peak, i.e., the ``flat-tail'' case, the model accommodates high-scale inflation compatible with tensor-to-scalar ratios of $r_\textrm{CMB} \lesssim 10^{-3}$ and reheating temperatures less than $\sim10^{14}$ GeV and greater than $\sim 10^4$~GeV ($10^8$ GeV) for the heaviest (lightest) PBHs in the asteroid mass window. This regime is particularly attractive as it allows for natural effective coupling scales $M_* \sim M_{\rm Pl}$ and remains a target for next-generation CMB B-mode experiments. Conversely, scenarios with sharper spectral decays ($n_{\rm exit} \to -4$), representing the extremely steep limit for single-field inflation, retain a strong ``memory'' of the USR velocity enhancement. To prevent baryon overproduction, the inflationary energy scale must be suppressed. Our results show that for $n_{\rm exit} = -4$, the tensor-to-scalar ratio is forced to extremely low values ($r_\textrm{CMB} \ll 10^{-12}$) and the reheating temperature to $T_{\rm reh} \lesssim 10^{11}$ GeV, with the same lower bounds as the flat-tail case of $\sim 10^4$~GeV ($10^8$ GeV) for the heaviest (lightest) PBHs. While invisible to CMB probes, this regime points toward small-field inflation scenarios, such as D-brane inflation (e.g., brane-antibrane annihilation), K\"ahler moduli inflation, or running mass models. In these theories, inherently small field excursions and low inflationary energy scales result in highly suppressed tensor amplitudes (vanishingly small $r_{\rm CMB}$), while the baryogenesis constraint selects this regime, leaving the induced SGWB as the sole discriminatory signature.

GW astronomy breaks the degeneracy between these two regimes. The large scalar perturbations required for PBH DM inevitably source a SGWB in the mHz--Hz frequency band. We find that both LISA and DECIGO will detect this background with exceptional significance, predicting signal-to-noise ratios of $\text{SNR} \sim 10^3 - 10^6$ and $\sim 10^4 - 10^8$, respectively, across the asteroid-mass window. Importantly, the ET can act as a discriminator for the inflationary dynamics. Our calculations show that ET is sensitive to the broadband high-frequency tail of the flat-tail scenario ($\text{SNR} \sim 10^5$) but loses sensitivity to the sharp cutoff characteristic of the steep-tail scenario. Consequently, a detection in LISA or DECIGO combined with a non-detection in ET would strongly favor a steep spectral decay ($n_{\rm exit} \lesssim -4$), indirectly probing the reheating history and favoring the low-scale inflationary nature of the baryogenesis mechanism.

This framework elevates USR inflation from a useful scenario for invoking PBH DM to a unifying physical mechanism connecting the smallest and largest observable inflationary scales. The direct correlation established here between the DM fraction, the baryon asymmetry, and the GW signal offers a striking signature. If spontaneous baryogenesis is indeed the origin of the matter-antimatter asymmetry, the presence of asteroid-mass PBHs and their associated GW background may provide compelling evidence.

%%%%%%%%%%%%%%%%%%%%%%%%%%%%%%%%%%%%%%%%%%%%%%%%
\section*{Acknowledgements}
%%%%%%%%%%%%%%%%%%%%%%%%%%%%%%%%%%%%%%%%%%%%%%%%
SB would like to thank Evan McDonough, Gabriele Franciolini and Guillem Dom\`enech for helpful feedback on the manuscript. SB is supported by the Science and Technology Facilities Council (STFC) under grant ST/X000753/1.

%%%%%%%%%%%%%%%%%%%%%%%%%%%%%%%%%%%%%%%%%%%%%%%%
\begin{widetext}
\appendix

%%%%%%%%%%%%%
\section{Scalar curvature power spectrum during ultra-slow-roll}
\label{app:USR}

In this work we model the enhancement of small-scale curvature perturbations by a brief phase of USR inflation. For completeness, we summarize the single-field origin of the characteristic superhorizon growth and the associated effective scaling of the power spectrum. See Refs.~\cite{Bassett:2005xm,Kinney:2005vj,Martin:2012pe,Byrnes:2018txb,Liu:2020oqe,Cole:2022xqc} for detailed discussions.

\paragraph{Equations of motion.}
For a canonical single inflaton, the comoving curvature perturbation ${\cal R}$ obeys the exact equation of motion \cite{Bassett:2005xm}
\beq
\frac{1}{a^3 \epsilon}\frac{d}{dt}\!\left(a^3 \epsilon \dot{\cal R}_k\right) + \frac{k^2}{a^2}\,{\cal R}_k = 0 ,
\label{eq:eomR_single}
\eeq
where $a(t)$ is the scale factor, and
\beq
\epsilon \equiv -\frac{\dot H}{H^2} = \frac{\dot{\phi}^{\,2}}{2H^2 M_{\rm Pl}^2}\,.
\eeq
Equivalently, in conformal time $\tau$ one may write the Mukhanov--Sasaki equation for $u_k \equiv z {\cal R}_k$ with
$z \equiv a \sqrt{2\epsilon}\, M_{\rm Pl}$,
\beq
u_k'' + \left(k^2 - \frac{z''}{z}\right)u_k = 0 ,
\label{eq:MS}
\eeq
where primes denote derivatives with respect to $\tau$.

\paragraph{Superhorizon solution and the non-attractor mode.}
On superhorizon scales ($k \ll aH$), Eq.~\eqref{eq:eomR_single} implies
\beq
\dot{\cal R}_k \simeq \frac{C_k}{a^3 \epsilon}\,,
\qquad
{\cal R}_k(t) \simeq C_{1k} + C_{2k}\int^t \frac{dt'}{a^3(t')\,\epsilon(t')}\,,
\label{eq:R_superhorizon}
\eeq
with constants $C_{1k},C_{2k}$ fixed by matching to the subhorizon evolution. During ordinary slow roll, $\epsilon$ varies slowly and the second mode decays as $a^{-3}$, so ${\cal R}_k$ is conserved outside the Hubble radius. During USR, however, $\epsilon$ falls rapidly and this second mode can become a growing mode.

\paragraph{USR background scaling and $a^3$ growth.}
In the idealized USR limit the inflaton evolves on an approximately flat portion of the potential, $V_{,\phi}\simeq 0$, so the Klein--Gordon equation reduces to
\beq
\ddot{\phi}+3H\dot{\phi}\simeq 0
\qquad \Rightarrow \qquad
\dot{\phi}\propto a^{-3},
\label{eq:USR_phidot_scaling_app}
\eeq
where we have used $H\simeq{\rm const}$ over the short duration of the USR phase. It follows that
\beq
\epsilon \equiv \frac{\dot{\phi}^{\,2}}{2H^2 M_{\rm Pl}^2}\propto a^{-6}.
\label{eq:USR_epsilon_scaling_app}
\eeq
Substituting $\epsilon\propto a^{-6}$ into the superhorizon solution~\eqref{eq:R_superhorizon} yields
\beq
\int^t \frac{dt'}{a^3(t')\,\epsilon(t')}
\;\propto\;
\int^t dt'\,a^3(t')
\;\propto\;
a^{3}(t),
\label{eq:USR_a3_growth_app}
\eeq
so that the would-be ``decaying'' mode becomes a growing mode during USR,
\beq
{\cal R}_k(t)\simeq C_{1k}+C_{2k}\,a^{3}(t)
\qquad (k\ll aH).
\label{eq:R_a3_growth_app}
\eeq
This parametric growth underlies the amplification of ${\cal P}_{\cal R}\propto |{\cal R}_k|^2$ in non-attractor phases and is equivalent to the familiar scaling ${\cal P}_{\cal R}\propto 1/\epsilon$ during the USR epoch.

\paragraph{Characterizing the net growth and the $k^4$ rise.}
Modes only grow while the background remains sufficiently non-attractor. A convenient way to quantify the net superhorizon enhancement is to introduce the (Hubble) second slow-roll parameter \footnote{In the main text we define $\eta \equiv \dot\epsilon/(H\epsilon)$, which yields $\eta\simeq -6$ in the ideal USR limit. For convenience, here, we use the alternative (Hubble) parameter
$\eta_H \equiv -\ddot\phi/(H\dot\phi)$, for which USR gives $\eta_H\simeq 3$. For $\epsilon\ll 1$ the two are related simply by $\eta = -2\eta_H + 2\epsilon \simeq -2\eta_H$, so the condition for superhorizon growth $\eta_H>3/2$ is equivalent to $\eta<-3$ in our convention.}
\beq
\eta_H \equiv -\frac{\ddot{\phi}}{H\dot{\phi}}\,,
\label{eq:etaH_def_app}
\eeq
for which the ideal USR solution~\eqref{eq:USR_phidot_scaling_app} gives $\eta_H\simeq 3$. In terms of $\eta_H$, superhorizon growth occurs when $\eta_H > 3/2$ \cite{Kinney:2005vj,Martin:2012pe,Liu:2020oqe,Cole:2022xqc}. One may therefore define the integrated departure from the threshold as \cite{Liu:2020oqe}
\beq
{\cal U} \equiv \int_{t_s}^{t_e} dt \left(\eta_H(t) - \frac{3}{2}\right) H(t)
\simeq \left(\bar{\eta}_H - \frac{3}{2}\right) N_{\rm usr}\,,
\label{eq:Udef_single}
\eeq
where $t_s$ and $t_e$ are defined by $\eta_H=3/2$ at the beginning and end of USR, $N_{\rm usr}$ is the USR duration in e-folds, and $\bar{\eta}_H$ is the average of $\eta_H$ over the USR phase.

Modes that exit the Hubble radius before the onset of USR can still be amplified if they are not too far outside at $t_s$. The smallest wavenumber that experiences net growth is \cite{Liu:2020oqe}
\beq
k_{\rm min} = k_s \exp\!\left(-{\cal U}\right)\,,
\qquad
k_s \equiv a(t_s)H(t_s)\,,
\label{eq:kmin_single}
\eeq
where $k_s$ is the scale exiting at the onset of USR. For modes in the range $k_{\rm min} \le k \le k_s$ that exit slightly before USR begins, the power spectrum acquires the characteristic effective scaling \cite{Liu:2020oqe} (see also Refs.~\cite{Martin:2012pe,Byrnes:2018txb,Cole:2022xqc})
\beq
{\cal P}_{\cal R}(k)
\simeq
\left(\frac{k}{k_s}\right)^4
e^{4{\cal U}}\,
{\cal P}_{\cal R}^{\rm SR}(k)\,,
\qquad
k_{\rm min} \le k \le k_s .
\label{eq:PR_k4}
\eeq
The $k^4$ behavior reflects the matching between the mode evolution prior to USR and the subsequent superhorizon growth of the non-attractor mode during USR. In the idealized limit $\eta_H\simeq 3$ one has ${\cal U}\simeq (3/2)N_{\rm usr}$ and the net amplification scales as $e^{6N_{\rm usr}}$, consistent with the parametric growth ${\cal P}_{\cal R}\propto 1/\epsilon$ during USR \cite{Kinney:2005vj,Martin:2012pe}.

\paragraph{Connection to the main text.}
In the main text we capture this physics phenomenologically by modeling the enhanced spectrum on small scales with a broken power-law form shown in Eq.~\eqref{eq:Pzeta_BPL_master} and an effective USR growth index (typically close to $n_{\rm USR}\simeq 4$), which reproduces the characteristic scaling in Eq.~\eqref{eq:PR_k4} (where $k_\textrm{on}$ plays the role of $k_\textrm{min}$ and the peak scale $k_\textrm{USR}$ plays the role of $k_s$ up to $\mathcal{O}(1)$ differences pertaining to how the onset/end of USR is defined in a smooth transition) over the range of modes amplified near the onset of USR. We do not attempt to incorporate additional constraints from non-Gaussian statistics of the perturbations, which can be relevant in some USR realizations and for PBH-related observables. We refer the reader to Refs.~\cite{Gow:2020bzo,Gow:2022jfb} for discussions of such effects.

\section{Thermal Baryon Asymmetry and Freeze--Out Condition}
\label{app:ThermalDerivation}

\subsection{Derivation of the Effective Chemical Potential}
\label{sec:chemicalpotential}
We begin by establishing the connection between the derivative coupling of the scalar field $\phi$ and the thermodynamic properties of the plasma. The interaction Lagrangian is given by
\begin{equation}
\mathcal{L}_{\rm int} = \frac{1}{M_*} (\partial_\mu \phi) j^\mu_B = \frac{1}{M_*} \dot{\phi} n_B - \frac{1}{M_*} (\nabla \phi) \cdot \mathbf{j}_B,
\end{equation}
where $j^\mu_B = (n_B, \mathbf{j}_B)$ is the baryon number current. In a spatially homogeneous background, $\nabla \phi = 0$, so the interaction reduces to
\begin{equation}
\mathcal{L}_{\rm int} = \frac{\dot{\phi}}{M_*} n_B.
\end{equation}
From the perspective of the statistical mechanics of the plasma, the Hamiltonian density is shifted by $\mathcal{H} \to \mathcal{H} - \mathcal{L}_{\rm int}$. Since the grand canonical partition function involves the operator $\hat{H} - \mu_B \hat{N}_B$, we can identify the coefficient of the number density operator $n_B$ as an effective chemical potential
\begin{equation}
\mu_B(t) = \frac{\dot{\phi}(t)}{M_*}.
\label{eq:chemicalpotential}
\end{equation}
This effective potential biases the energy levels of particles versus antiparticles, leading to a net asymmetry in thermal equilibrium.

\subsection{Baryon Density Calculation}

Consider a relativistic fermionic species $i$ carrying baryon number $B_i$ and its antiparticle $\bar{i}$ with baryon number $-B_i$.
In the presence of a baryon-number chemical potential $\mu_B$, the equilibrium phase space distribution functions are
\begin{equation}
f(p, \mu_i) = \frac{1}{\exp[(E_p - \mu_i)/T] + 1}, \qquad
\bar{f}(p, \mu_i) = \frac{1}{\exp[(E_p + \mu_i)/T] + 1},
\end{equation}
where $E_p = \sqrt{p^2 + m^2}$ and $\mu_i = B_i \mu_B$.
The net number density is
\begin{equation}
n_i - n_{\bar{i}} = g_i \int \frac{d^3p}{(2\pi)^3} \left[ f(p, \mu_i) - \bar{f}(p, \mu_i) \right],
\end{equation}
where $g_i$ accounts for internal degrees of freedom (spin and color).

Assuming $\mu_i \ll T$ (and the relativistic limit $m \ll T$), we can Taylor expand to first order in $\mu_i$
\begin{equation}
\frac{1}{e^{(E-\mu)/T} + 1} - \frac{1}{e^{(E+\mu)/T} + 1}
\approx
\frac{\partial f}{\partial \mu}\bigg|_{\mu=0} (2\mu_i)
=
-\frac{\partial f}{\partial E} (2\mu_i)
=
\frac{2\mu_i}{T}\frac{e^{E/T}}{(e^{E/T}+1)^2}.
\end{equation}
Substituting this expansion into the integral yields
\begin{equation}
n_i - n_{\bar{i}} \simeq \frac{g_i}{2\pi^2} \frac{2\mu_i}{T}
\int_0^\infty dp \, p^2 \frac{e^{p/T}}{(e^{p/T}+1)^2}.
\end{equation}
Using $x=p/T$ and the standard integral
\begin{equation}
\int_0^\infty dx \frac{x^2 e^x}{(e^x + 1)^2} = \frac{\pi^2}{6},
\end{equation}
we obtain
\begin{equation}
n_i - n_{\bar{i}} \simeq \frac{g_i}{6}\,\mu_i\,T^2.
\end{equation}
The net baryon number density is therefore
\begin{equation}
n_B = \sum_i B_i\,(n_i-n_{\bar i})
\simeq \frac{\mu_B T^2}{6}\sum_i g_i B_i^2
\equiv \frac{g_B}{6}\,\mu_B T^2,
\label{eq:app_nB_derived}
\end{equation}
which defines $g_B \equiv \sum_i g_i B_i^2$ as used in the main text.
This confirms the result stated in Eq.~\eqref{eq:nBs_basic} of the main text.

\subsection{Entropy Density and Final Yield}

The total entropy density of the relativistic plasma is dominated by radiation and is given by
\begin{equation}
s = \frac{2\pi^2}{45} g_*(T) T^3.
\end{equation}
Combining this with Eq.~\eqref{eq:app_nB_derived} gives the equilibrium baryon-to-entropy ratio:
\begin{equation}
\left( \frac{n_B}{s} \right)_{\rm eq} = \frac{g_B/6 \, \mu_B T^2}{2\pi^2/45 \, g_* T^3} = \frac{15 g_B}{4\pi^2 g_*} \frac{\mu_B}{T}.
\end{equation}
Finally, we address the freeze-out condition. The evolution of the asymmetry is governed by the Boltzmann equation
\begin{equation}
\frac{d n_B}{dt} + 3Hn_B = -\Gamma_{\Delta B} (n_B - n_B^{\rm eq}).
\end{equation}
We assume a sudden freeze-out approximation. At high temperatures ($T > T_D$), the interaction rate $\Gamma_{\Delta B} \gg H$, enforcing $n_B = n_B^{\rm eq}$. At $T_D$, defined by $\Gamma_{\Delta B}(T_D) \simeq H(T_D)$, the interactions decouple. For $T < T_D$, the baryon number is conserved in a comoving volume, meaning $n_B \propto a^{-3}$. Since entropy also scales as $s \propto a^{-3}$ (assuming no subsequent entropy production), the ratio $n_B/s$ becomes constant
\begin{equation}
\frac{n_B}{s}\bigg|_{\rm today} = \frac{n_B}{s}\bigg|_{T_D} = \frac{15 g_B}{4\pi^2 g_*} \frac{\dot{\phi}(T_D)}{M_* T_D}.
\end{equation}

\end{widetext}

\bibliography{refs.bib}

\begin{thebibliography}{71}%
\makeatletter
\providecommand \@ifxundefined [1]{%
 \@ifx{#1\undefined}
}%
\providecommand \@ifnum [1]{%
 \ifnum #1\expandafter \@firstoftwo
 \else \expandafter \@secondoftwo
 \fi
}%
\providecommand \@ifx [1]{%
 \ifx #1\expandafter \@firstoftwo
 \else \expandafter \@secondoftwo
 \fi
}%
\providecommand \natexlab [1]{#1}%
\providecommand \enquote  [1]{``#1''}%
\providecommand \bibnamefont  [1]{#1}%
\providecommand \bibfnamefont [1]{#1}%
\providecommand \citenamefont [1]{#1}%
\providecommand \href@noop [0]{\@secondoftwo}%
\providecommand \href [0]{\begingroup \@sanitize@url \@href}%
\providecommand \@href[1]{\@@startlink{#1}\@@href}%
\providecommand \@@href[1]{\endgroup#1\@@endlink}%
\providecommand \@sanitize@url [0]{\catcode `\\12\catcode `\$12\catcode `\&12\catcode `\#12\catcode `\^12\catcode `\_12\catcode `\%12\relax}%
\providecommand \@@startlink[1]{}%
\providecommand \@@endlink[0]{}%
\providecommand \url  [0]{\begingroup\@sanitize@url \@url }%
\providecommand \@url [1]{\endgroup\@href {#1}{\urlprefix }}%
\providecommand \urlprefix  [0]{URL }%
\providecommand \Eprint [0]{\href }%
\providecommand \doibase [0]{http://dx.doi.org/}%
\providecommand \selectlanguage [0]{\@gobble}%
\providecommand \bibinfo  [0]{\@secondoftwo}%
\providecommand \bibfield  [0]{\@secondoftwo}%
\providecommand \translation [1]{[#1]}%
\providecommand \BibitemOpen [0]{}%
\providecommand \bibitemStop [0]{}%
\providecommand \bibitemNoStop [0]{.\EOS\space}%
\providecommand \EOS [0]{\spacefactor3000\relax}%
\providecommand \BibitemShut  [1]{\csname bibitem#1\endcsname}%
\let\auto@bib@innerbib\@empty
%</preamble>
\bibitem [{\citenamefont {Carr}\ \emph {et~al.}(2021)\citenamefont {Carr}, \citenamefont {Kohri}, \citenamefont {Sendouda},\ and\ \citenamefont {Yokoyama}}]{Carr:2020gox}%
  \BibitemOpen
  \bibfield  {author} {\bibinfo {author} {\bibfnamefont {Bernard}\ \bibnamefont {Carr}}, \bibinfo {author} {\bibfnamefont {Kazunori}\ \bibnamefont {Kohri}}, \bibinfo {author} {\bibfnamefont {Yuuiti}\ \bibnamefont {Sendouda}}, \ and\ \bibinfo {author} {\bibfnamefont {Jun'ichi}\ \bibnamefont {Yokoyama}},\ }\bibfield  {title} {\enquote {\bibinfo {title} {{Constraints on primordial black holes}},}\ }\href {\doibase 10.1088/1361-6633/ac1e31} {\bibfield  {journal} {\bibinfo  {journal} {Rept. Prog. Phys.}\ }\textbf {\bibinfo {volume} {84}},\ \bibinfo {pages} {116902} (\bibinfo {year} {2021})},\ \Eprint {http://arxiv.org/abs/2002.12778} {arXiv:2002.12778 [astro-ph.CO]} \BibitemShut {NoStop}%
\bibitem [{\citenamefont {Carr}\ and\ \citenamefont {Kuhnel}(2025)}]{Carr:2025kdk}%
  \BibitemOpen
  \bibfield  {author} {\bibinfo {author} {\bibfnamefont {Bernard}\ \bibnamefont {Carr}}\ and\ \bibinfo {author} {\bibfnamefont {Florian}\ \bibnamefont {Kuhnel}},\ }\bibfield  {title} {\enquote {\bibinfo {title} {{Primordial Black Holes}},}\ }\href@noop {} {\  (\bibinfo {year} {2025})},\ \Eprint {http://arxiv.org/abs/2502.15279} {arXiv:2502.15279 [astro-ph.CO]} \BibitemShut {NoStop}%
\bibitem [{\citenamefont {Garcia-Bellido}\ and\ \citenamefont {Ruiz~Morales}(2017)}]{Garcia-Bellido:2017mdw}%
  \BibitemOpen
  \bibfield  {author} {\bibinfo {author} {\bibfnamefont {Juan}\ \bibnamefont {Garcia-Bellido}}\ and\ \bibinfo {author} {\bibfnamefont {Ester}\ \bibnamefont {Ruiz~Morales}},\ }\bibfield  {title} {\enquote {\bibinfo {title} {{Primordial black holes from single field models of inflation}},}\ }\href {\doibase 10.1016/j.dark.2017.09.007} {\bibfield  {journal} {\bibinfo  {journal} {Phys. Dark Univ.}\ }\textbf {\bibinfo {volume} {18}},\ \bibinfo {pages} {47--54} (\bibinfo {year} {2017})},\ \Eprint {http://arxiv.org/abs/1702.03901} {arXiv:1702.03901 [astro-ph.CO]} \BibitemShut {NoStop}%
\bibitem [{\citenamefont {Ezquiaga}\ \emph {et~al.}(2018)\citenamefont {Ezquiaga}, \citenamefont {Garcia-Bellido},\ and\ \citenamefont {Ruiz~Morales}}]{Ezquiaga:2017fvi}%
  \BibitemOpen
  \bibfield  {author} {\bibinfo {author} {\bibfnamefont {Jose~Maria}\ \bibnamefont {Ezquiaga}}, \bibinfo {author} {\bibfnamefont {Juan}\ \bibnamefont {Garcia-Bellido}}, \ and\ \bibinfo {author} {\bibfnamefont {Ester}\ \bibnamefont {Ruiz~Morales}},\ }\bibfield  {title} {\enquote {\bibinfo {title} {{Primordial Black Hole production in Critical Higgs Inflation}},}\ }\href {\doibase 10.1016/j.physletb.2017.11.039} {\bibfield  {journal} {\bibinfo  {journal} {Phys. Lett. B}\ }\textbf {\bibinfo {volume} {776}},\ \bibinfo {pages} {345--349} (\bibinfo {year} {2018})},\ \Eprint {http://arxiv.org/abs/1705.04861} {arXiv:1705.04861 [astro-ph.CO]} \BibitemShut {NoStop}%
\bibitem [{\citenamefont {Germani}\ and\ \citenamefont {Prokopec}(2017)}]{Germani:2017bcs}%
  \BibitemOpen
  \bibfield  {author} {\bibinfo {author} {\bibfnamefont {Cristiano}\ \bibnamefont {Germani}}\ and\ \bibinfo {author} {\bibfnamefont {Tomislav}\ \bibnamefont {Prokopec}},\ }\bibfield  {title} {\enquote {\bibinfo {title} {{On primordial black holes from an inflection point}},}\ }\href {\doibase 10.1016/j.dark.2017.09.001} {\bibfield  {journal} {\bibinfo  {journal} {Phys. Dark Univ.}\ }\textbf {\bibinfo {volume} {18}},\ \bibinfo {pages} {6--10} (\bibinfo {year} {2017})},\ \Eprint {http://arxiv.org/abs/1706.04226} {arXiv:1706.04226 [astro-ph.CO]} \BibitemShut {NoStop}%
\bibitem [{\citenamefont {Kannike}\ \emph {et~al.}(2017)\citenamefont {Kannike}, \citenamefont {Marzola}, \citenamefont {Raidal},\ and\ \citenamefont {Veerm\"ae}}]{Kannike:2017bxn}%
  \BibitemOpen
  \bibfield  {author} {\bibinfo {author} {\bibfnamefont {Kristjan}\ \bibnamefont {Kannike}}, \bibinfo {author} {\bibfnamefont {Luca}\ \bibnamefont {Marzola}}, \bibinfo {author} {\bibfnamefont {Martti}\ \bibnamefont {Raidal}}, \ and\ \bibinfo {author} {\bibfnamefont {Hardi}\ \bibnamefont {Veerm\"ae}},\ }\bibfield  {title} {\enquote {\bibinfo {title} {{Single Field Double Inflation and Primordial Black Holes}},}\ }\href {\doibase 10.1088/1475-7516/2017/09/020} {\bibfield  {journal} {\bibinfo  {journal} {JCAP}\ }\textbf {\bibinfo {volume} {09}},\ \bibinfo {pages} {020} (\bibinfo {year} {2017})},\ \Eprint {http://arxiv.org/abs/1705.06225} {arXiv:1705.06225 [astro-ph.CO]} \BibitemShut {NoStop}%
\bibitem [{\citenamefont {Motohashi}\ and\ \citenamefont {Hu}(2017)}]{Motohashi:2017kbs}%
  \BibitemOpen
  \bibfield  {author} {\bibinfo {author} {\bibfnamefont {Hayato}\ \bibnamefont {Motohashi}}\ and\ \bibinfo {author} {\bibfnamefont {Wayne}\ \bibnamefont {Hu}},\ }\bibfield  {title} {\enquote {\bibinfo {title} {{Primordial Black Holes and Slow-Roll Violation}},}\ }\href {\doibase 10.1103/PhysRevD.96.063503} {\bibfield  {journal} {\bibinfo  {journal} {Phys. Rev. D}\ }\textbf {\bibinfo {volume} {96}},\ \bibinfo {pages} {063503} (\bibinfo {year} {2017})},\ \Eprint {http://arxiv.org/abs/1706.06784} {arXiv:1706.06784 [astro-ph.CO]} \BibitemShut {NoStop}%
\bibitem [{\citenamefont {Di}\ and\ \citenamefont {Gong}(2018)}]{Di:2017ndc}%
  \BibitemOpen
  \bibfield  {author} {\bibinfo {author} {\bibfnamefont {Haoran}\ \bibnamefont {Di}}\ and\ \bibinfo {author} {\bibfnamefont {Yungui}\ \bibnamefont {Gong}},\ }\bibfield  {title} {\enquote {\bibinfo {title} {{Primordial black holes and second order gravitational waves from ultra-slow-roll inflation}},}\ }\href {\doibase 10.1088/1475-7516/2018/07/007} {\bibfield  {journal} {\bibinfo  {journal} {JCAP}\ }\textbf {\bibinfo {volume} {07}},\ \bibinfo {pages} {007} (\bibinfo {year} {2018})},\ \Eprint {http://arxiv.org/abs/1707.09578} {arXiv:1707.09578 [astro-ph.CO]} \BibitemShut {NoStop}%
\bibitem [{\citenamefont {Ballesteros}\ and\ \citenamefont {Taoso}(2018)}]{Ballesteros:2017fsr}%
  \BibitemOpen
  \bibfield  {author} {\bibinfo {author} {\bibfnamefont {Guillermo}\ \bibnamefont {Ballesteros}}\ and\ \bibinfo {author} {\bibfnamefont {Marco}\ \bibnamefont {Taoso}},\ }\bibfield  {title} {\enquote {\bibinfo {title} {{Primordial black hole dark matter from single field inflation}},}\ }\href {\doibase 10.1103/PhysRevD.97.023501} {\bibfield  {journal} {\bibinfo  {journal} {Phys. Rev. D}\ }\textbf {\bibinfo {volume} {97}},\ \bibinfo {pages} {023501} (\bibinfo {year} {2018})},\ \Eprint {http://arxiv.org/abs/1709.05565} {arXiv:1709.05565 [hep-ph]} \BibitemShut {NoStop}%
\bibitem [{\citenamefont {Pattison}\ \emph {et~al.}(2017)\citenamefont {Pattison}, \citenamefont {Vennin}, \citenamefont {Assadullahi},\ and\ \citenamefont {Wands}}]{Pattison:2017mbe}%
  \BibitemOpen
  \bibfield  {author} {\bibinfo {author} {\bibfnamefont {Chris}\ \bibnamefont {Pattison}}, \bibinfo {author} {\bibfnamefont {Vincent}\ \bibnamefont {Vennin}}, \bibinfo {author} {\bibfnamefont {Hooshyar}\ \bibnamefont {Assadullahi}}, \ and\ \bibinfo {author} {\bibfnamefont {David}\ \bibnamefont {Wands}},\ }\bibfield  {title} {\enquote {\bibinfo {title} {{Quantum diffusion during inflation and primordial black holes}},}\ }\href {\doibase 10.1088/1475-7516/2017/10/046} {\bibfield  {journal} {\bibinfo  {journal} {JCAP}\ }\textbf {\bibinfo {volume} {10}},\ \bibinfo {pages} {046} (\bibinfo {year} {2017})},\ \Eprint {http://arxiv.org/abs/1707.00537} {arXiv:1707.00537 [hep-th]} \BibitemShut {NoStop}%
\bibitem [{\citenamefont {Passaglia}\ \emph {et~al.}(2019)\citenamefont {Passaglia}, \citenamefont {Hu},\ and\ \citenamefont {Motohashi}}]{Passaglia:2018ixg}%
  \BibitemOpen
  \bibfield  {author} {\bibinfo {author} {\bibfnamefont {Samuel}\ \bibnamefont {Passaglia}}, \bibinfo {author} {\bibfnamefont {Wayne}\ \bibnamefont {Hu}}, \ and\ \bibinfo {author} {\bibfnamefont {Hayato}\ \bibnamefont {Motohashi}},\ }\bibfield  {title} {\enquote {\bibinfo {title} {{Primordial black holes and local non-Gaussianity in canonical inflation}},}\ }\href {\doibase 10.1103/PhysRevD.99.043536} {\bibfield  {journal} {\bibinfo  {journal} {Phys. Rev. D}\ }\textbf {\bibinfo {volume} {99}},\ \bibinfo {pages} {043536} (\bibinfo {year} {2019})},\ \Eprint {http://arxiv.org/abs/1812.08243} {arXiv:1812.08243 [astro-ph.CO]} \BibitemShut {NoStop}%
\bibitem [{\citenamefont {Biagetti}\ \emph {et~al.}(2018)\citenamefont {Biagetti}, \citenamefont {Franciolini}, \citenamefont {Kehagias},\ and\ \citenamefont {Riotto}}]{Biagetti:2018pjj}%
  \BibitemOpen
  \bibfield  {author} {\bibinfo {author} {\bibfnamefont {Matteo}\ \bibnamefont {Biagetti}}, \bibinfo {author} {\bibfnamefont {Gabriele}\ \bibnamefont {Franciolini}}, \bibinfo {author} {\bibfnamefont {Alex}\ \bibnamefont {Kehagias}}, \ and\ \bibinfo {author} {\bibfnamefont {Antonio}\ \bibnamefont {Riotto}},\ }\bibfield  {title} {\enquote {\bibinfo {title} {{Primordial Black Holes from Inflation and Quantum Diffusion}},}\ }\href {\doibase 10.1088/1475-7516/2018/07/032} {\bibfield  {journal} {\bibinfo  {journal} {JCAP}\ }\textbf {\bibinfo {volume} {07}},\ \bibinfo {pages} {032} (\bibinfo {year} {2018})},\ \Eprint {http://arxiv.org/abs/1804.07124} {arXiv:1804.07124 [astro-ph.CO]} \BibitemShut {NoStop}%
\bibitem [{\citenamefont {Byrnes}\ \emph {et~al.}(2019)\citenamefont {Byrnes}, \citenamefont {Cole},\ and\ \citenamefont {Patil}}]{Byrnes:2018txb}%
  \BibitemOpen
  \bibfield  {author} {\bibinfo {author} {\bibfnamefont {Christian~T.}\ \bibnamefont {Byrnes}}, \bibinfo {author} {\bibfnamefont {Philippa~S.}\ \bibnamefont {Cole}}, \ and\ \bibinfo {author} {\bibfnamefont {Subodh~P.}\ \bibnamefont {Patil}},\ }\bibfield  {title} {\enquote {\bibinfo {title} {{Steepest growth of the power spectrum and primordial black holes}},}\ }\href {\doibase 10.1088/1475-7516/2019/06/028} {\bibfield  {journal} {\bibinfo  {journal} {JCAP}\ }\textbf {\bibinfo {volume} {06}},\ \bibinfo {pages} {028} (\bibinfo {year} {2019})},\ \Eprint {http://arxiv.org/abs/1811.11158} {arXiv:1811.11158 [astro-ph.CO]} \BibitemShut {NoStop}%
\bibitem [{\citenamefont {Carrilho}\ \emph {et~al.}(2019)\citenamefont {Carrilho}, \citenamefont {Malik},\ and\ \citenamefont {Mulryne}}]{Carrilho:2019oqg}%
  \BibitemOpen
  \bibfield  {author} {\bibinfo {author} {\bibfnamefont {Pedro}\ \bibnamefont {Carrilho}}, \bibinfo {author} {\bibfnamefont {Karim~A.}\ \bibnamefont {Malik}}, \ and\ \bibinfo {author} {\bibfnamefont {David~J.}\ \bibnamefont {Mulryne}},\ }\bibfield  {title} {\enquote {\bibinfo {title} {{Dissecting the growth of the power spectrum for primordial black holes}},}\ }\href {\doibase 10.1103/PhysRevD.100.103529} {\bibfield  {journal} {\bibinfo  {journal} {Phys. Rev. D}\ }\textbf {\bibinfo {volume} {100}},\ \bibinfo {pages} {103529} (\bibinfo {year} {2019})},\ \Eprint {http://arxiv.org/abs/1907.05237} {arXiv:1907.05237 [astro-ph.CO]} \BibitemShut {NoStop}%
\bibitem [{\citenamefont {Figueroa}\ \emph {et~al.}(2021)\citenamefont {Figueroa}, \citenamefont {Raatikainen}, \citenamefont {Rasanen},\ and\ \citenamefont {Tomberg}}]{Figueroa:2020jkf}%
  \BibitemOpen
  \bibfield  {author} {\bibinfo {author} {\bibfnamefont {Daniel~G.}\ \bibnamefont {Figueroa}}, \bibinfo {author} {\bibfnamefont {Sami}\ \bibnamefont {Raatikainen}}, \bibinfo {author} {\bibfnamefont {Syksy}\ \bibnamefont {Rasanen}}, \ and\ \bibinfo {author} {\bibfnamefont {Eemeli}\ \bibnamefont {Tomberg}},\ }\bibfield  {title} {\enquote {\bibinfo {title} {{Non-Gaussian Tail of the Curvature Perturbation in Stochastic Ultraslow-Roll Inflation: Implications for Primordial Black Hole Production}},}\ }\href {\doibase 10.1103/PhysRevLett.127.101302} {\bibfield  {journal} {\bibinfo  {journal} {Phys. Rev. Lett.}\ }\textbf {\bibinfo {volume} {127}},\ \bibinfo {pages} {101302} (\bibinfo {year} {2021})},\ \Eprint {http://arxiv.org/abs/2012.06551} {arXiv:2012.06551 [astro-ph.CO]} \BibitemShut {NoStop}%
\bibitem [{\citenamefont {Inomata}\ \emph {et~al.}(2022)\citenamefont {Inomata}, \citenamefont {McDonough},\ and\ \citenamefont {Hu}}]{Inomata:2021tpx}%
  \BibitemOpen
  \bibfield  {author} {\bibinfo {author} {\bibfnamefont {Keisuke}\ \bibnamefont {Inomata}}, \bibinfo {author} {\bibfnamefont {Evan}\ \bibnamefont {McDonough}}, \ and\ \bibinfo {author} {\bibfnamefont {Wayne}\ \bibnamefont {Hu}},\ }\bibfield  {title} {\enquote {\bibinfo {title} {{Amplification of primordial perturbations from the rise or fall of the inflaton}},}\ }\href {\doibase 10.1088/1475-7516/2022/02/031} {\bibfield  {journal} {\bibinfo  {journal} {JCAP}\ }\textbf {\bibinfo {volume} {02}},\ \bibinfo {pages} {031} (\bibinfo {year} {2022})},\ \Eprint {http://arxiv.org/abs/2110.14641} {arXiv:2110.14641 [astro-ph.CO]} \BibitemShut {NoStop}%
\bibitem [{\citenamefont {Inomata}\ \emph {et~al.}(2021)\citenamefont {Inomata}, \citenamefont {McDonough},\ and\ \citenamefont {Hu}}]{Inomata:2021uqj}%
  \BibitemOpen
  \bibfield  {author} {\bibinfo {author} {\bibfnamefont {Keisuke}\ \bibnamefont {Inomata}}, \bibinfo {author} {\bibfnamefont {Evan}\ \bibnamefont {McDonough}}, \ and\ \bibinfo {author} {\bibfnamefont {Wayne}\ \bibnamefont {Hu}},\ }\bibfield  {title} {\enquote {\bibinfo {title} {{Primordial black holes arise when the inflaton falls}},}\ }\href {\doibase 10.1103/PhysRevD.104.123553} {\bibfield  {journal} {\bibinfo  {journal} {Phys. Rev. D}\ }\textbf {\bibinfo {volume} {104}},\ \bibinfo {pages} {123553} (\bibinfo {year} {2021})},\ \Eprint {http://arxiv.org/abs/2104.03972} {arXiv:2104.03972 [astro-ph.CO]} \BibitemShut {NoStop}%
\bibitem [{\citenamefont {Pattison}\ \emph {et~al.}(2021)\citenamefont {Pattison}, \citenamefont {Vennin}, \citenamefont {Wands},\ and\ \citenamefont {Assadullahi}}]{Pattison:2021oen}%
  \BibitemOpen
  \bibfield  {author} {\bibinfo {author} {\bibfnamefont {Chris}\ \bibnamefont {Pattison}}, \bibinfo {author} {\bibfnamefont {Vincent}\ \bibnamefont {Vennin}}, \bibinfo {author} {\bibfnamefont {David}\ \bibnamefont {Wands}}, \ and\ \bibinfo {author} {\bibfnamefont {Hooshyar}\ \bibnamefont {Assadullahi}},\ }\bibfield  {title} {\enquote {\bibinfo {title} {{Ultra-slow-roll inflation with quantum diffusion}},}\ }\href {\doibase 10.1088/1475-7516/2021/04/080} {\bibfield  {journal} {\bibinfo  {journal} {JCAP}\ }\textbf {\bibinfo {volume} {04}},\ \bibinfo {pages} {080} (\bibinfo {year} {2021})},\ \Eprint {http://arxiv.org/abs/2101.05741} {arXiv:2101.05741 [astro-ph.CO]} \BibitemShut {NoStop}%
\bibitem [{\citenamefont {Balaji}\ \emph {et~al.}(2022{\natexlab{a}})\citenamefont {Balaji}, \citenamefont {Silk},\ and\ \citenamefont {Wu}}]{Balaji:2022rsy}%
  \BibitemOpen
  \bibfield  {author} {\bibinfo {author} {\bibfnamefont {Shyam}\ \bibnamefont {Balaji}}, \bibinfo {author} {\bibfnamefont {Joseph}\ \bibnamefont {Silk}}, \ and\ \bibinfo {author} {\bibfnamefont {Yi-Peng}\ \bibnamefont {Wu}},\ }\bibfield  {title} {\enquote {\bibinfo {title} {{Induced gravitational waves from the cosmic coincidence}},}\ }\href {\doibase 10.1088/1475-7516/2022/06/008} {\bibfield  {journal} {\bibinfo  {journal} {JCAP}\ }\textbf {\bibinfo {volume} {06}},\ \bibinfo {pages} {008} (\bibinfo {year} {2022}{\natexlab{a}})},\ \Eprint {http://arxiv.org/abs/2202.00700} {arXiv:2202.00700 [astro-ph.CO]} \BibitemShut {NoStop}%
\bibitem [{\citenamefont {Balaji}\ \emph {et~al.}(2022{\natexlab{b}})\citenamefont {Balaji}, \citenamefont {Ragavendra}, \citenamefont {Sethi}, \citenamefont {Silk},\ and\ \citenamefont {Sriramkumar}}]{Balaji:2022zur}%
  \BibitemOpen
  \bibfield  {author} {\bibinfo {author} {\bibfnamefont {Shyam}\ \bibnamefont {Balaji}}, \bibinfo {author} {\bibfnamefont {H.~V.}\ \bibnamefont {Ragavendra}}, \bibinfo {author} {\bibfnamefont {Shiv~K.}\ \bibnamefont {Sethi}}, \bibinfo {author} {\bibfnamefont {Joseph}\ \bibnamefont {Silk}}, \ and\ \bibinfo {author} {\bibfnamefont {L.}~\bibnamefont {Sriramkumar}},\ }\bibfield  {title} {\enquote {\bibinfo {title} {{Observing Nulling of Primordial Correlations via the 21-cm Signal}},}\ }\href {\doibase 10.1103/PhysRevLett.129.261301} {\bibfield  {journal} {\bibinfo  {journal} {Phys. Rev. Lett.}\ }\textbf {\bibinfo {volume} {129}},\ \bibinfo {pages} {261301} (\bibinfo {year} {2022}{\natexlab{b}})},\ \Eprint {http://arxiv.org/abs/2206.06386} {arXiv:2206.06386 [astro-ph.CO]} \BibitemShut {NoStop}%
\bibitem [{\citenamefont {Balaji}\ \emph {et~al.}(2022{\natexlab{c}})\citenamefont {Balaji}, \citenamefont {Domenech},\ and\ \citenamefont {Silk}}]{Balaji:2022dbi}%
  \BibitemOpen
  \bibfield  {author} {\bibinfo {author} {\bibfnamefont {Shyam}\ \bibnamefont {Balaji}}, \bibinfo {author} {\bibfnamefont {Guillem}\ \bibnamefont {Domenech}}, \ and\ \bibinfo {author} {\bibfnamefont {Joseph}\ \bibnamefont {Silk}},\ }\bibfield  {title} {\enquote {\bibinfo {title} {{Induced gravitational waves from slow-roll inflation after an enhancing phase}},}\ }\href {\doibase 10.1088/1475-7516/2022/09/016} {\bibfield  {journal} {\bibinfo  {journal} {JCAP}\ }\textbf {\bibinfo {volume} {09}},\ \bibinfo {pages} {016} (\bibinfo {year} {2022}{\natexlab{c}})},\ \Eprint {http://arxiv.org/abs/2205.01696} {arXiv:2205.01696 [astro-ph.CO]} \BibitemShut {NoStop}%
\bibitem [{\citenamefont {Kawai}\ and\ \citenamefont {Kim}(2023)}]{Kawai:2022emp}%
  \BibitemOpen
  \bibfield  {author} {\bibinfo {author} {\bibfnamefont {Shinsuke}\ \bibnamefont {Kawai}}\ and\ \bibinfo {author} {\bibfnamefont {Jinsu}\ \bibnamefont {Kim}},\ }\bibfield  {title} {\enquote {\bibinfo {title} {{Primordial black holes and gravitational waves from nonminimally coupled supergravity inflation}},}\ }\href {\doibase 10.1103/PhysRevD.107.043523} {\bibfield  {journal} {\bibinfo  {journal} {Phys. Rev. D}\ }\textbf {\bibinfo {volume} {107}},\ \bibinfo {pages} {043523} (\bibinfo {year} {2023})},\ \Eprint {http://arxiv.org/abs/2209.15343} {arXiv:2209.15343 [astro-ph.CO]} \BibitemShut {NoStop}%
\bibitem [{\citenamefont {Karam}\ \emph {et~al.}(2023)\citenamefont {Karam}, \citenamefont {Koivunen}, \citenamefont {Tomberg}, \citenamefont {Vaskonen},\ and\ \citenamefont {Veerm\"ae}}]{Karam:2022nym}%
  \BibitemOpen
  \bibfield  {author} {\bibinfo {author} {\bibfnamefont {Alexandros}\ \bibnamefont {Karam}}, \bibinfo {author} {\bibfnamefont {Niko}\ \bibnamefont {Koivunen}}, \bibinfo {author} {\bibfnamefont {Eemeli}\ \bibnamefont {Tomberg}}, \bibinfo {author} {\bibfnamefont {Ville}\ \bibnamefont {Vaskonen}}, \ and\ \bibinfo {author} {\bibfnamefont {Hardi}\ \bibnamefont {Veerm\"ae}},\ }\bibfield  {title} {\enquote {\bibinfo {title} {{Anatomy of single-field inflationary models for primordial black holes}},}\ }\href {\doibase 10.1088/1475-7516/2023/03/013} {\bibfield  {journal} {\bibinfo  {journal} {JCAP}\ }\textbf {\bibinfo {volume} {03}},\ \bibinfo {pages} {013} (\bibinfo {year} {2023})},\ \Eprint {http://arxiv.org/abs/2205.13540} {arXiv:2205.13540 [astro-ph.CO]} \BibitemShut {NoStop}%
\bibitem [{\citenamefont {Pi}\ and\ \citenamefont {Sasaki}(2022)}]{Pi:2022ysn}%
  \BibitemOpen
  \bibfield  {author} {\bibinfo {author} {\bibfnamefont {Shi}\ \bibnamefont {Pi}}\ and\ \bibinfo {author} {\bibfnamefont {Misao}\ \bibnamefont {Sasaki}},\ }\bibfield  {title} {\enquote {\bibinfo {title} {{Logarithmic Duality of the Curvature Perturbation}},}\ }\href@noop {} {\  (\bibinfo {year} {2022})},\ \Eprint {http://arxiv.org/abs/2211.13932} {arXiv:2211.13932 [astro-ph.CO]} \BibitemShut {NoStop}%
\bibitem [{\citenamefont {Qin}\ \emph {et~al.}(2023)\citenamefont {Qin}, \citenamefont {Geller}, \citenamefont {Balaji}, \citenamefont {McDonough},\ and\ \citenamefont {Kaiser}}]{Qin:2023lgo}%
  \BibitemOpen
  \bibfield  {author} {\bibinfo {author} {\bibfnamefont {Wenzer}\ \bibnamefont {Qin}}, \bibinfo {author} {\bibfnamefont {Sarah~R.}\ \bibnamefont {Geller}}, \bibinfo {author} {\bibfnamefont {Shyam}\ \bibnamefont {Balaji}}, \bibinfo {author} {\bibfnamefont {Evan}\ \bibnamefont {McDonough}}, \ and\ \bibinfo {author} {\bibfnamefont {David~I.}\ \bibnamefont {Kaiser}},\ }\bibfield  {title} {\enquote {\bibinfo {title} {{Planck constraints and gravitational wave forecasts for primordial black hole dark matter seeded by multifield inflation}},}\ }\href {\doibase 10.1103/PhysRevD.108.043508} {\bibfield  {journal} {\bibinfo  {journal} {Phys. Rev. D}\ }\textbf {\bibinfo {volume} {108}},\ \bibinfo {pages} {043508} (\bibinfo {year} {2023})},\ \Eprint {http://arxiv.org/abs/2303.02168} {arXiv:2303.02168 [astro-ph.CO]} \BibitemShut {NoStop}%
\bibitem [{\citenamefont {Geller}\ \emph {et~al.}(2022)\citenamefont {Geller}, \citenamefont {Qin}, \citenamefont {McDonough},\ and\ \citenamefont {Kaiser}}]{Geller:2022nkr}%
  \BibitemOpen
  \bibfield  {author} {\bibinfo {author} {\bibfnamefont {Sarah~R.}\ \bibnamefont {Geller}}, \bibinfo {author} {\bibfnamefont {Wenzer}\ \bibnamefont {Qin}}, \bibinfo {author} {\bibfnamefont {Evan}\ \bibnamefont {McDonough}}, \ and\ \bibinfo {author} {\bibfnamefont {David~I.}\ \bibnamefont {Kaiser}},\ }\bibfield  {title} {\enquote {\bibinfo {title} {{Primordial black holes from multifield inflation with nonminimal couplings}},}\ }\href {\doibase 10.1103/PhysRevD.106.063535} {\bibfield  {journal} {\bibinfo  {journal} {Phys. Rev. D}\ }\textbf {\bibinfo {volume} {106}},\ \bibinfo {pages} {063535} (\bibinfo {year} {2022})},\ \Eprint {http://arxiv.org/abs/2205.04471} {arXiv:2205.04471 [hep-th]} \BibitemShut {NoStop}%
\bibitem [{\citenamefont {Lorenzoni}\ \emph {et~al.}(2025)\citenamefont {Lorenzoni}, \citenamefont {Geller}, \citenamefont {Ireland}, \citenamefont {Kaiser}, \citenamefont {McDonough},\ and\ \citenamefont {Wittmeier}}]{Lorenzoni:2025gni}%
  \BibitemOpen
  \bibfield  {author} {\bibinfo {author} {\bibfnamefont {Dario~L.}\ \bibnamefont {Lorenzoni}}, \bibinfo {author} {\bibfnamefont {Sarah~R.}\ \bibnamefont {Geller}}, \bibinfo {author} {\bibfnamefont {Zachary}\ \bibnamefont {Ireland}}, \bibinfo {author} {\bibfnamefont {David~I.}\ \bibnamefont {Kaiser}}, \bibinfo {author} {\bibfnamefont {Evan}\ \bibnamefont {McDonough}}, \ and\ \bibinfo {author} {\bibfnamefont {Kyle~A.}\ \bibnamefont {Wittmeier}},\ }\bibfield  {title} {\enquote {\bibinfo {title} {{Light Scalar Fields Foster Production of Primordial Black Holes}},}\ }\href@noop {} {\  (\bibinfo {year} {2025})},\ \Eprint {http://arxiv.org/abs/2504.13251} {arXiv:2504.13251 [astro-ph.CO]} \BibitemShut {NoStop}%
\bibitem [{\citenamefont {Kinney}(2005)}]{Kinney:2005vj}%
  \BibitemOpen
  \bibfield  {author} {\bibinfo {author} {\bibfnamefont {William~H.}\ \bibnamefont {Kinney}},\ }\bibfield  {title} {\enquote {\bibinfo {title} {{Horizon crossing and inflation with large eta}},}\ }\href {\doibase 10.1103/PhysRevD.72.023515} {\bibfield  {journal} {\bibinfo  {journal} {Phys. Rev. D}\ }\textbf {\bibinfo {volume} {72}},\ \bibinfo {pages} {023515} (\bibinfo {year} {2005})},\ \Eprint {http://arxiv.org/abs/gr-qc/0503017} {arXiv:gr-qc/0503017} \BibitemShut {NoStop}%
\bibitem [{\citenamefont {Martin}\ \emph {et~al.}(2013)\citenamefont {Martin}, \citenamefont {Motohashi},\ and\ \citenamefont {Suyama}}]{Martin:2012pe}%
  \BibitemOpen
  \bibfield  {author} {\bibinfo {author} {\bibfnamefont {Jerome}\ \bibnamefont {Martin}}, \bibinfo {author} {\bibfnamefont {Hayato}\ \bibnamefont {Motohashi}}, \ and\ \bibinfo {author} {\bibfnamefont {Teruaki}\ \bibnamefont {Suyama}},\ }\bibfield  {title} {\enquote {\bibinfo {title} {{Ultra Slow-Roll Inflation and the non-Gaussianity Consistency Relation}},}\ }\href {\doibase 10.1103/PhysRevD.87.023514} {\bibfield  {journal} {\bibinfo  {journal} {Phys. Rev. D}\ }\textbf {\bibinfo {volume} {87}},\ \bibinfo {pages} {023514} (\bibinfo {year} {2013})},\ \Eprint {http://arxiv.org/abs/1211.0083} {arXiv:1211.0083 [astro-ph.CO]} \BibitemShut {NoStop}%
\bibitem [{\citenamefont {Liu}\ \emph {et~al.}(2020)\citenamefont {Liu}, \citenamefont {Guo},\ and\ \citenamefont {Cai}}]{Liu:2020oqe}%
  \BibitemOpen
  \bibfield  {author} {\bibinfo {author} {\bibfnamefont {Jing}\ \bibnamefont {Liu}}, \bibinfo {author} {\bibfnamefont {Zong-Kuan}\ \bibnamefont {Guo}}, \ and\ \bibinfo {author} {\bibfnamefont {Rong-Gen}\ \bibnamefont {Cai}},\ }\bibfield  {title} {\enquote {\bibinfo {title} {{Analytical approximation of the scalar spectrum in the ultraslow-roll inflationary models}},}\ }\href {\doibase 10.1103/PhysRevD.101.083535} {\bibfield  {journal} {\bibinfo  {journal} {Phys. Rev. D}\ }\textbf {\bibinfo {volume} {101}},\ \bibinfo {pages} {083535} (\bibinfo {year} {2020})},\ \Eprint {http://arxiv.org/abs/2003.02075} {arXiv:2003.02075 [astro-ph.CO]} \BibitemShut {NoStop}%
\bibitem [{\citenamefont {Cole}\ \emph {et~al.}(2022)\citenamefont {Cole}, \citenamefont {Gow}, \citenamefont {Byrnes},\ and\ \citenamefont {Patil}}]{Cole:2022xqc}%
  \BibitemOpen
  \bibfield  {author} {\bibinfo {author} {\bibfnamefont {Philippa~S.}\ \bibnamefont {Cole}}, \bibinfo {author} {\bibfnamefont {Andrew~D.}\ \bibnamefont {Gow}}, \bibinfo {author} {\bibfnamefont {Christian~T.}\ \bibnamefont {Byrnes}}, \ and\ \bibinfo {author} {\bibfnamefont {Subodh~P.}\ \bibnamefont {Patil}},\ }\bibfield  {title} {\enquote {\bibinfo {title} {{Steepest growth re-examined: repercussions for primordial black hole formation}},}\ }\href@noop {} {\  (\bibinfo {year} {2022})},\ \Eprint {http://arxiv.org/abs/2204.07573} {arXiv:2204.07573 [astro-ph.CO]} \BibitemShut {NoStop}%
\bibitem [{\citenamefont {Khlopov}(2010)}]{Khlopov:2008qy}%
  \BibitemOpen
  \bibfield  {author} {\bibinfo {author} {\bibfnamefont {Maxim~Yu.}\ \bibnamefont {Khlopov}},\ }\bibfield  {title} {\enquote {\bibinfo {title} {{Primordial Black Holes}},}\ }\href {\doibase 10.1088/1674-4527/10/6/001} {\bibfield  {journal} {\bibinfo  {journal} {Res. Astron. Astrophys.}\ }\textbf {\bibinfo {volume} {10}},\ \bibinfo {pages} {495--528} (\bibinfo {year} {2010})},\ \Eprint {http://arxiv.org/abs/0801.0116} {arXiv:0801.0116 [astro-ph]} \BibitemShut {NoStop}%
\bibitem [{\citenamefont {Sasaki}\ \emph {et~al.}(2018)\citenamefont {Sasaki}, \citenamefont {Suyama}, \citenamefont {Tanaka},\ and\ \citenamefont {Yokoyama}}]{Sasaki:2018dmp}%
  \BibitemOpen
  \bibfield  {author} {\bibinfo {author} {\bibfnamefont {Misao}\ \bibnamefont {Sasaki}}, \bibinfo {author} {\bibfnamefont {Teruaki}\ \bibnamefont {Suyama}}, \bibinfo {author} {\bibfnamefont {Takahiro}\ \bibnamefont {Tanaka}}, \ and\ \bibinfo {author} {\bibfnamefont {Shuichiro}\ \bibnamefont {Yokoyama}},\ }\bibfield  {title} {\enquote {\bibinfo {title} {{Primordial black holes\textemdash{}perspectives in gravitational wave astronomy}},}\ }\href {\doibase 10.1088/1361-6382/aaa7b4} {\bibfield  {journal} {\bibinfo  {journal} {Class. Quant. Grav.}\ }\textbf {\bibinfo {volume} {35}},\ \bibinfo {pages} {063001} (\bibinfo {year} {2018})},\ \Eprint {http://arxiv.org/abs/1801.05235} {arXiv:1801.05235 [astro-ph.CO]} \BibitemShut {NoStop}%
\bibitem [{\citenamefont {{Carr}}\ and\ \citenamefont {{K\"{u}hnel}}(2020)}]{Carr:2020xqk}%
  \BibitemOpen
  \bibfield  {author} {\bibinfo {author} {\bibfnamefont {Bernard}\ \bibnamefont {{Carr}}}\ and\ \bibinfo {author} {\bibfnamefont {Florian}\ \bibnamefont {{K\"{u}hnel}}},\ }\bibfield  {title} {\enquote {\bibinfo {title} {{Primordial Black Holes as Dark Matter: Recent Developments}},}\ }\href {\doibase 10.1146/annurev-nucl-050520-125911} {\bibfield  {journal} {\bibinfo  {journal} {Ann. Rev. Nucl. Part. Sci.}\ }\textbf {\bibinfo {volume} {70}},\ \bibinfo {pages} {355--394} (\bibinfo {year} {2020})},\ \Eprint {http://arxiv.org/abs/2006.02838} {arXiv:2006.02838 [astro-ph.CO]} \BibitemShut {NoStop}%
\bibitem [{\citenamefont {Green}\ and\ \citenamefont {Kavanagh}(2021)}]{Green:2020jor}%
  \BibitemOpen
  \bibfield  {author} {\bibinfo {author} {\bibfnamefont {Anne~M.}\ \bibnamefont {Green}}\ and\ \bibinfo {author} {\bibfnamefont {Bradley~J.}\ \bibnamefont {Kavanagh}},\ }\bibfield  {title} {\enquote {\bibinfo {title} {{Primordial Black Holes as a dark matter candidate}},}\ }\href {\doibase 10.1088/1361-6471/abc534} {\bibfield  {journal} {\bibinfo  {journal} {J. Phys. G}\ }\textbf {\bibinfo {volume} {48}},\ \bibinfo {pages} {043001} (\bibinfo {year} {2021})},\ \Eprint {http://arxiv.org/abs/2007.10722} {arXiv:2007.10722 [astro-ph.CO]} \BibitemShut {NoStop}%
\bibitem [{\citenamefont {Escriv\`a}(2022)}]{Escriva:2021aeh}%
  \BibitemOpen
  \bibfield  {author} {\bibinfo {author} {\bibfnamefont {Albert}\ \bibnamefont {Escriv\`a}},\ }\bibfield  {title} {\enquote {\bibinfo {title} {{PBH Formation from Spherically Symmetric Hydrodynamical Perturbations: A Review}},}\ }\href {\doibase 10.3390/universe8020066} {\bibfield  {journal} {\bibinfo  {journal} {Universe}\ }\textbf {\bibinfo {volume} {8}},\ \bibinfo {pages} {66} (\bibinfo {year} {2022})},\ \Eprint {http://arxiv.org/abs/2111.12693} {arXiv:2111.12693 [gr-qc]} \BibitemShut {NoStop}%
\bibitem [{\citenamefont {Villanueva-Domingo}\ \emph {et~al.}(2021)\citenamefont {Villanueva-Domingo}, \citenamefont {Mena},\ and\ \citenamefont {Palomares-Ruiz}}]{Villanueva-Domingo:2021spv}%
  \BibitemOpen
  \bibfield  {author} {\bibinfo {author} {\bibfnamefont {Pablo}\ \bibnamefont {Villanueva-Domingo}}, \bibinfo {author} {\bibfnamefont {Olga}\ \bibnamefont {Mena}}, \ and\ \bibinfo {author} {\bibfnamefont {Sergio}\ \bibnamefont {Palomares-Ruiz}},\ }\bibfield  {title} {\enquote {\bibinfo {title} {{A brief review on primordial black holes as dark matter}},}\ }\href {\doibase 10.3389/fspas.2021.681084} {\bibfield  {journal} {\bibinfo  {journal} {Front. Astron. Space Sci.}\ }\textbf {\bibinfo {volume} {8}},\ \bibinfo {pages} {87} (\bibinfo {year} {2021})},\ \Eprint {http://arxiv.org/abs/2103.12087} {arXiv:2103.12087 [astro-ph.CO]} \BibitemShut {NoStop}%
\bibitem [{\citenamefont {Escriv\`a}\ \emph {et~al.}(2022)\citenamefont {Escriv\`a}, \citenamefont {Kuhnel},\ and\ \citenamefont {Tada}}]{Escriva:2022duf}%
  \BibitemOpen
  \bibfield  {author} {\bibinfo {author} {\bibfnamefont {Albert}\ \bibnamefont {Escriv\`a}}, \bibinfo {author} {\bibfnamefont {Florian}\ \bibnamefont {Kuhnel}}, \ and\ \bibinfo {author} {\bibfnamefont {Yuichiro}\ \bibnamefont {Tada}},\ }\bibfield  {title} {\enquote {\bibinfo {title} {{Primordial Black Holes}},}\ }\href@noop {} {\  (\bibinfo {year} {2022})},\ \Eprint {http://arxiv.org/abs/2211.05767} {arXiv:2211.05767 [astro-ph.CO]} \BibitemShut {NoStop}%
\bibitem [{\citenamefont {\"Ozsoy}\ and\ \citenamefont {Tasinato}(2023)}]{Ozsoy:2023ryl}%
  \BibitemOpen
  \bibfield  {author} {\bibinfo {author} {\bibfnamefont {Ogan}\ \bibnamefont {\"Ozsoy}}\ and\ \bibinfo {author} {\bibfnamefont {Gianmassimo}\ \bibnamefont {Tasinato}},\ }\bibfield  {title} {\enquote {\bibinfo {title} {{Inflation and Primordial Black Holes}},}\ }\href@noop {} {\  (\bibinfo {year} {2023})},\ \Eprint {http://arxiv.org/abs/2301.03600} {arXiv:2301.03600 [astro-ph.CO]} \BibitemShut {NoStop}%
\bibitem [{\citenamefont {Cohen}\ and\ \citenamefont {Kaplan}(1987)}]{Cohen:1987vi}%
  \BibitemOpen
  \bibfield  {author} {\bibinfo {author} {\bibfnamefont {Andrew~G.}\ \bibnamefont {Cohen}}\ and\ \bibinfo {author} {\bibfnamefont {David~B.}\ \bibnamefont {Kaplan}},\ }\bibfield  {title} {\enquote {\bibinfo {title} {{Thermodynamic Generation of the Baryon Asymmetry}},}\ }\href {\doibase 10.1016/0370-2693(87)91369-4} {\bibfield  {journal} {\bibinfo  {journal} {Phys. Lett. B}\ }\textbf {\bibinfo {volume} {199}},\ \bibinfo {pages} {251--258} (\bibinfo {year} {1987})}\BibitemShut {NoStop}%
\bibitem [{\citenamefont {Cohen}\ and\ \citenamefont {Kaplan}(1988)}]{Cohen:1988kt}%
  \BibitemOpen
  \bibfield  {author} {\bibinfo {author} {\bibfnamefont {Andrew~G.}\ \bibnamefont {Cohen}}\ and\ \bibinfo {author} {\bibfnamefont {David~B.}\ \bibnamefont {Kaplan}},\ }\bibfield  {title} {\enquote {\bibinfo {title} {{SPONTANEOUS BARYOGENESIS}},}\ }\href {\doibase 10.1016/0550-3213(88)90134-4} {\bibfield  {journal} {\bibinfo  {journal} {Nucl. Phys. B}\ }\textbf {\bibinfo {volume} {308}},\ \bibinfo {pages} {913--928} (\bibinfo {year} {1988})}\BibitemShut {NoStop}%
\bibitem [{\citenamefont {Akrami}\ \emph {et~al.}(2020)\citenamefont {Akrami} \emph {et~al.}}]{Planck:2018jri}%
  \BibitemOpen
  \bibfield  {author} {\bibinfo {author} {\bibfnamefont {Y.}~\bibnamefont {Akrami}} \emph {et~al.} (\bibinfo {collaboration} {Planck}),\ }\bibfield  {title} {\enquote {\bibinfo {title} {{Planck 2018 results. X. Constraints on inflation}},}\ }\href {\doibase 10.1051/0004-6361/201833887} {\bibfield  {journal} {\bibinfo  {journal} {Astron. Astrophys.}\ }\textbf {\bibinfo {volume} {641}},\ \bibinfo {pages} {A10} (\bibinfo {year} {2020})},\ \Eprint {http://arxiv.org/abs/1807.06211} {arXiv:1807.06211 [astro-ph.CO]} \BibitemShut {NoStop}%
\bibitem [{\citenamefont {Guzzetti}\ \emph {et~al.}(2016)\citenamefont {Guzzetti}, \citenamefont {Bartolo}, \citenamefont {Liguori},\ and\ \citenamefont {Matarrese}}]{Guzzetti:2016mkm}%
  \BibitemOpen
  \bibfield  {author} {\bibinfo {author} {\bibfnamefont {M.~C.}\ \bibnamefont {Guzzetti}}, \bibinfo {author} {\bibfnamefont {N.}~\bibnamefont {Bartolo}}, \bibinfo {author} {\bibfnamefont {M.}~\bibnamefont {Liguori}}, \ and\ \bibinfo {author} {\bibfnamefont {S.}~\bibnamefont {Matarrese}},\ }\bibfield  {title} {\enquote {\bibinfo {title} {{Gravitational waves from inflation}},}\ }\href {\doibase 10.1393/ncr/i2016-10127-1} {\bibfield  {journal} {\bibinfo  {journal} {Riv. Nuovo Cim.}\ }\textbf {\bibinfo {volume} {39}},\ \bibinfo {pages} {399--495} (\bibinfo {year} {2016})},\ \Eprint {http://arxiv.org/abs/1605.01615} {arXiv:1605.01615 [astro-ph.CO]} \BibitemShut {NoStop}%
\bibitem [{\citenamefont {Caprini}\ and\ \citenamefont {Figueroa}(2018)}]{Caprini:2018mtu}%
  \BibitemOpen
  \bibfield  {author} {\bibinfo {author} {\bibfnamefont {Chiara}\ \bibnamefont {Caprini}}\ and\ \bibinfo {author} {\bibfnamefont {Daniel~G.}\ \bibnamefont {Figueroa}},\ }\bibfield  {title} {\enquote {\bibinfo {title} {{Cosmological Backgrounds of Gravitational Waves}},}\ }\href {\doibase 10.1088/1361-6382/aac608} {\bibfield  {journal} {\bibinfo  {journal} {Class. Quant. Grav.}\ }\textbf {\bibinfo {volume} {35}},\ \bibinfo {pages} {163001} (\bibinfo {year} {2018})},\ \Eprint {http://arxiv.org/abs/1801.04268} {arXiv:1801.04268 [astro-ph.CO]} \BibitemShut {NoStop}%
\bibitem [{\citenamefont {Dom\`enech}(2021)}]{Domenech:2021ztg}%
  \BibitemOpen
  \bibfield  {author} {\bibinfo {author} {\bibfnamefont {Guillem}\ \bibnamefont {Dom\`enech}},\ }\bibfield  {title} {\enquote {\bibinfo {title} {{Scalar Induced Gravitational Waves Review}},}\ }\href {\doibase 10.3390/universe7110398} {\bibfield  {journal} {\bibinfo  {journal} {Universe}\ }\textbf {\bibinfo {volume} {7}},\ \bibinfo {pages} {398} (\bibinfo {year} {2021})},\ \Eprint {http://arxiv.org/abs/2109.01398} {arXiv:2109.01398 [gr-qc]} \BibitemShut {NoStop}%
\bibitem [{\citenamefont {Barausse}\ \emph {et~al.}(2020)\citenamefont {Barausse} \emph {et~al.}}]{Barausse:2020rsu}%
  \BibitemOpen
  \bibfield  {author} {\bibinfo {author} {\bibfnamefont {Enrico}\ \bibnamefont {Barausse}} \emph {et~al.},\ }\bibfield  {title} {\enquote {\bibinfo {title} {{Prospects for Fundamental Physics with LISA}},}\ }\href {\doibase 10.1007/s10714-020-02691-1} {\bibfield  {journal} {\bibinfo  {journal} {Gen. Rel. Grav.}\ }\textbf {\bibinfo {volume} {52}},\ \bibinfo {pages} {81} (\bibinfo {year} {2020})},\ \Eprint {http://arxiv.org/abs/2001.09793} {arXiv:2001.09793 [gr-qc]} \BibitemShut {NoStop}%
\bibitem [{\citenamefont {Yuan}\ and\ \citenamefont {Huang}(2021)}]{Yuan:2021qgz}%
  \BibitemOpen
  \bibfield  {author} {\bibinfo {author} {\bibfnamefont {Chen}\ \bibnamefont {Yuan}}\ and\ \bibinfo {author} {\bibfnamefont {Qing-Guo}\ \bibnamefont {Huang}},\ }\bibfield  {title} {\enquote {\bibinfo {title} {{A topic review on probing primordial black hole dark matter with scalar induced gravitational waves}},}\ }\href@noop {} {\  (\bibinfo {year} {2021})},\ \Eprint {http://arxiv.org/abs/2103.04739} {arXiv:2103.04739 [astro-ph.GA]} \BibitemShut {NoStop}%
\bibitem [{\citenamefont {Kawamura}\ \emph {et~al.}(2021)\citenamefont {Kawamura} \emph {et~al.}}]{Kawamura:2020pcg}%
  \BibitemOpen
  \bibfield  {author} {\bibinfo {author} {\bibfnamefont {Seiji}\ \bibnamefont {Kawamura}} \emph {et~al.},\ }\bibfield  {title} {\enquote {\bibinfo {title} {{Current status of space gravitational wave antenna DECIGO and B-DECIGO}},}\ }\href {\doibase 10.1093/ptep/ptab019} {\bibfield  {journal} {\bibinfo  {journal} {PTEP}\ }\textbf {\bibinfo {volume} {2021}},\ \bibinfo {pages} {05A105} (\bibinfo {year} {2021})},\ \Eprint {http://arxiv.org/abs/2006.13545} {arXiv:2006.13545 [gr-qc]} \BibitemShut {NoStop}%
\bibitem [{\citenamefont {Maggiore}\ \emph {et~al.}(2020)\citenamefont {Maggiore} \emph {et~al.}}]{Maggiore:2019uih}%
  \BibitemOpen
  \bibfield  {author} {\bibinfo {author} {\bibfnamefont {Michele}\ \bibnamefont {Maggiore}} \emph {et~al.},\ }\bibfield  {title} {\enquote {\bibinfo {title} {{Science Case for the Einstein Telescope}},}\ }\href {\doibase 10.1088/1475-7516/2020/03/050} {\bibfield  {journal} {\bibinfo  {journal} {JCAP}\ }\textbf {\bibinfo {volume} {03}},\ \bibinfo {pages} {050} (\bibinfo {year} {2020})},\ \Eprint {http://arxiv.org/abs/1912.02622} {arXiv:1912.02622 [astro-ph.CO]} \BibitemShut {NoStop}%
\bibitem [{\citenamefont {De~Simone}\ and\ \citenamefont {Kobayashi}(2016)}]{DeSimone:2016ofp}%
  \BibitemOpen
  \bibfield  {author} {\bibinfo {author} {\bibfnamefont {Andrea}\ \bibnamefont {De~Simone}}\ and\ \bibinfo {author} {\bibfnamefont {Takeshi}\ \bibnamefont {Kobayashi}},\ }\bibfield  {title} {\enquote {\bibinfo {title} {{Cosmological Aspects of Spontaneous Baryogenesis}},}\ }\href {\doibase 10.1088/1475-7516/2016/08/052} {\bibfield  {journal} {\bibinfo  {journal} {JCAP}\ }\textbf {\bibinfo {volume} {08}},\ \bibinfo {pages} {052} (\bibinfo {year} {2016})},\ \Eprint {http://arxiv.org/abs/1605.00670} {arXiv:1605.00670 [hep-ph]} \BibitemShut {NoStop}%
\bibitem [{\citenamefont {Inomata}\ \emph {et~al.}(2018)\citenamefont {Inomata}, \citenamefont {Kawasaki}, \citenamefont {Kusenko},\ and\ \citenamefont {Yang}}]{Inomata:2018htm}%
  \BibitemOpen
  \bibfield  {author} {\bibinfo {author} {\bibfnamefont {Keisuke}\ \bibnamefont {Inomata}}, \bibinfo {author} {\bibfnamefont {Masahiro}\ \bibnamefont {Kawasaki}}, \bibinfo {author} {\bibfnamefont {Alexander}\ \bibnamefont {Kusenko}}, \ and\ \bibinfo {author} {\bibfnamefont {Louis}\ \bibnamefont {Yang}},\ }\bibfield  {title} {\enquote {\bibinfo {title} {{Big Bang Nucleosynthesis Constraint on Baryonic Isocurvature Perturbations}},}\ }\href {\doibase 10.1088/1475-7516/2018/12/003} {\bibfield  {journal} {\bibinfo  {journal} {JCAP}\ }\textbf {\bibinfo {volume} {12}},\ \bibinfo {pages} {003} (\bibinfo {year} {2018})},\ \Eprint {http://arxiv.org/abs/1806.00123} {arXiv:1806.00123 [astro-ph.CO]} \BibitemShut {NoStop}%
\bibitem [{\citenamefont {Kawasaki}\ and\ \citenamefont {Nakatsuka}(2019)}]{Kawasaki:2019mbl}%
  \BibitemOpen
  \bibfield  {author} {\bibinfo {author} {\bibfnamefont {Masahiro}\ \bibnamefont {Kawasaki}}\ and\ \bibinfo {author} {\bibfnamefont {Hiromasa}\ \bibnamefont {Nakatsuka}},\ }\bibfield  {title} {\enquote {\bibinfo {title} {{Effect of nonlinearity between density and curvature perturbations on the primordial black hole formation}},}\ }\href {\doibase 10.1103/PhysRevD.99.123501} {\bibfield  {journal} {\bibinfo  {journal} {Phys. Rev. D}\ }\textbf {\bibinfo {volume} {99}},\ \bibinfo {pages} {123501} (\bibinfo {year} {2019})},\ \Eprint {http://arxiv.org/abs/1903.02994} {arXiv:1903.02994 [astro-ph.CO]} \BibitemShut {NoStop}%
\bibitem [{\citenamefont {Aghanim}\ \emph {et~al.}(2020)\citenamefont {Aghanim} \emph {et~al.}}]{Planck:2018vyg}%
  \BibitemOpen
  \bibfield  {author} {\bibinfo {author} {\bibfnamefont {N.}~\bibnamefont {Aghanim}} \emph {et~al.} (\bibinfo {collaboration} {Planck}),\ }\bibfield  {title} {\enquote {\bibinfo {title} {{Planck 2018 results. VI. Cosmological parameters}},}\ }\href {\doibase 10.1051/0004-6361/201833910} {\bibfield  {journal} {\bibinfo  {journal} {Astron. Astrophys.}\ }\textbf {\bibinfo {volume} {641}},\ \bibinfo {pages} {A6} (\bibinfo {year} {2020})},\ \bibinfo {note} {[Erratum: Astron.Astrophys. 652, C4 (2021)]},\ \Eprint {http://arxiv.org/abs/1807.06209} {arXiv:1807.06209 [astro-ph.CO]} \BibitemShut {NoStop}%
\bibitem [{\citenamefont {Carr}(1975)}]{Carr:1975qj}%
  \BibitemOpen
  \bibfield  {author} {\bibinfo {author} {\bibfnamefont {Bernard~J.}\ \bibnamefont {Carr}},\ }\bibfield  {title} {\enquote {\bibinfo {title} {{The Primordial black hole mass spectrum}},}\ }\href {\doibase 10.1086/153853} {\bibfield  {journal} {\bibinfo  {journal} {Astrophys. J.}\ }\textbf {\bibinfo {volume} {201}},\ \bibinfo {pages} {1--19} (\bibinfo {year} {1975})}\BibitemShut {NoStop}%
\bibitem [{\citenamefont {Musco}\ and\ \citenamefont {Miller}(2013)}]{Musco:2012au}%
  \BibitemOpen
  \bibfield  {author} {\bibinfo {author} {\bibfnamefont {Ilia}\ \bibnamefont {Musco}}\ and\ \bibinfo {author} {\bibfnamefont {John~C.}\ \bibnamefont {Miller}},\ }\bibfield  {title} {\enquote {\bibinfo {title} {{Primordial black hole formation in the early universe: critical behaviour and self-similarity}},}\ }\href {\doibase 10.1088/0264-9381/30/14/145009} {\bibfield  {journal} {\bibinfo  {journal} {Class. Quant. Grav.}\ }\textbf {\bibinfo {volume} {30}},\ \bibinfo {pages} {145009} (\bibinfo {year} {2013})},\ \Eprint {http://arxiv.org/abs/1201.2379} {arXiv:1201.2379 [gr-qc]} \BibitemShut {NoStop}%
\bibitem [{\citenamefont {Harada}\ \emph {et~al.}(2013)\citenamefont {Harada}, \citenamefont {Yoo},\ and\ \citenamefont {Kohri}}]{Harada:2013epa}%
  \BibitemOpen
  \bibfield  {author} {\bibinfo {author} {\bibfnamefont {Tomohiro}\ \bibnamefont {Harada}}, \bibinfo {author} {\bibfnamefont {Chul-Moon}\ \bibnamefont {Yoo}}, \ and\ \bibinfo {author} {\bibfnamefont {Kazunori}\ \bibnamefont {Kohri}},\ }\bibfield  {title} {\enquote {\bibinfo {title} {{Threshold of primordial black hole formation}},}\ }\href {\doibase 10.1103/PhysRevD.88.084051} {\bibfield  {journal} {\bibinfo  {journal} {Phys. Rev. D}\ }\textbf {\bibinfo {volume} {88}},\ \bibinfo {pages} {084051} (\bibinfo {year} {2013})},\ \bibinfo {note} {[Erratum: Phys.Rev.D 89, 029903 (2014)]},\ \Eprint {http://arxiv.org/abs/1309.4201} {arXiv:1309.4201 [astro-ph.CO]} \BibitemShut {NoStop}%
\bibitem [{\citenamefont {Gow}\ \emph {et~al.}(2021)\citenamefont {Gow}, \citenamefont {Byrnes}, \citenamefont {Cole},\ and\ \citenamefont {Young}}]{Gow:2020bzo}%
  \BibitemOpen
  \bibfield  {author} {\bibinfo {author} {\bibfnamefont {Andrew~D.}\ \bibnamefont {Gow}}, \bibinfo {author} {\bibfnamefont {Christian~T.}\ \bibnamefont {Byrnes}}, \bibinfo {author} {\bibfnamefont {Philippa~S.}\ \bibnamefont {Cole}}, \ and\ \bibinfo {author} {\bibfnamefont {Sam}\ \bibnamefont {Young}},\ }\bibfield  {title} {\enquote {\bibinfo {title} {{The power spectrum on small scales: Robust constraints and comparing PBH methodologies}},}\ }\href {\doibase 10.1088/1475-7516/2021/02/002} {\bibfield  {journal} {\bibinfo  {journal} {JCAP}\ }\textbf {\bibinfo {volume} {02}},\ \bibinfo {pages} {002} (\bibinfo {year} {2021})},\ \Eprint {http://arxiv.org/abs/2008.03289} {arXiv:2008.03289 [astro-ph.CO]} \BibitemShut {NoStop}%
\bibitem [{\citenamefont {Gow}\ \emph {et~al.}(2022)\citenamefont {Gow}, \citenamefont {Assadullahi}, \citenamefont {Jackson}, \citenamefont {Koyama}, \citenamefont {Vennin},\ and\ \citenamefont {Wands}}]{Gow:2022jfb}%
  \BibitemOpen
  \bibfield  {author} {\bibinfo {author} {\bibfnamefont {Andrew~D.}\ \bibnamefont {Gow}}, \bibinfo {author} {\bibfnamefont {Hooshyar}\ \bibnamefont {Assadullahi}}, \bibinfo {author} {\bibfnamefont {Joseph H.~P.}\ \bibnamefont {Jackson}}, \bibinfo {author} {\bibfnamefont {Kazuya}\ \bibnamefont {Koyama}}, \bibinfo {author} {\bibfnamefont {Vincent}\ \bibnamefont {Vennin}}, \ and\ \bibinfo {author} {\bibfnamefont {David}\ \bibnamefont {Wands}},\ }\bibfield  {title} {\enquote {\bibinfo {title} {{Non-perturbative non-Gaussianity and primordial black holes}},}\ }\href@noop {} {\  (\bibinfo {year} {2022})},\ \Eprint {http://arxiv.org/abs/2211.08348} {arXiv:2211.08348 [astro-ph.CO]} \BibitemShut {NoStop}%
\bibitem [{\citenamefont {Musco}(2019)}]{Musco:2018rwt}%
  \BibitemOpen
  \bibfield  {author} {\bibinfo {author} {\bibfnamefont {Ilia}\ \bibnamefont {Musco}},\ }\bibfield  {title} {\enquote {\bibinfo {title} {{Threshold for primordial black holes: Dependence on the shape of the cosmological perturbations}},}\ }\href {\doibase 10.1103/PhysRevD.100.123524} {\bibfield  {journal} {\bibinfo  {journal} {Phys. Rev. D}\ }\textbf {\bibinfo {volume} {100}},\ \bibinfo {pages} {123524} (\bibinfo {year} {2019})},\ \Eprint {http://arxiv.org/abs/1809.02127} {arXiv:1809.02127 [gr-qc]} \BibitemShut {NoStop}%
\bibitem [{\citenamefont {Ferrante}\ \emph {et~al.}(2023)\citenamefont {Ferrante}, \citenamefont {Franciolini}, \citenamefont {Iovino},\ and\ \citenamefont {Urbano}}]{Ferrante:2022mui}%
  \BibitemOpen
  \bibfield  {author} {\bibinfo {author} {\bibfnamefont {Giacomo}\ \bibnamefont {Ferrante}}, \bibinfo {author} {\bibfnamefont {Gabriele}\ \bibnamefont {Franciolini}}, \bibinfo {author} {\bibfnamefont {Antonio}\ \bibnamefont {Iovino}, \bibfnamefont {Junior.}}, \ and\ \bibinfo {author} {\bibfnamefont {Alfredo}\ \bibnamefont {Urbano}},\ }\bibfield  {title} {\enquote {\bibinfo {title} {{Primordial non-Gaussianity up to all orders: Theoretical aspects and implications for primordial black hole models}},}\ }\href {\doibase 10.1103/PhysRevD.107.043520} {\bibfield  {journal} {\bibinfo  {journal} {Phys. Rev. D}\ }\textbf {\bibinfo {volume} {107}},\ \bibinfo {pages} {043520} (\bibinfo {year} {2023})},\ \Eprint {http://arxiv.org/abs/2211.01728} {arXiv:2211.01728 [astro-ph.CO]} \BibitemShut {NoStop}%
\bibitem [{\citenamefont {De~la Torre~Luque}\ \emph {et~al.}(2024)\citenamefont {De~la Torre~Luque}, \citenamefont {Koechler},\ and\ \citenamefont {Balaji}}]{DelaTorreLuque:2024qms}%
  \BibitemOpen
  \bibfield  {author} {\bibinfo {author} {\bibfnamefont {Pedro}\ \bibnamefont {De~la Torre~Luque}}, \bibinfo {author} {\bibfnamefont {Jordan}\ \bibnamefont {Koechler}}, \ and\ \bibinfo {author} {\bibfnamefont {Shyam}\ \bibnamefont {Balaji}},\ }\bibfield  {title} {\enquote {\bibinfo {title} {{Refining Galactic primordial black hole evaporation constraints}},}\ }\href {\doibase 10.1103/PhysRevD.110.123022} {\bibfield  {journal} {\bibinfo  {journal} {Phys. Rev. D}\ }\textbf {\bibinfo {volume} {110}},\ \bibinfo {pages} {123022} (\bibinfo {year} {2024})},\ \bibinfo {note} {[Erratum: Phys.Rev.D 112, 109904 (2025)]},\ \Eprint {http://arxiv.org/abs/2406.11949} {arXiv:2406.11949 [astro-ph.HE]} \BibitemShut {NoStop}%
\bibitem [{\citenamefont {Ananda}\ \emph {et~al.}(2007)\citenamefont {Ananda}, \citenamefont {Clarkson},\ and\ \citenamefont {Wands}}]{Ananda:2006af}%
  \BibitemOpen
  \bibfield  {author} {\bibinfo {author} {\bibfnamefont {Kishore~N.}\ \bibnamefont {Ananda}}, \bibinfo {author} {\bibfnamefont {Chris}\ \bibnamefont {Clarkson}}, \ and\ \bibinfo {author} {\bibfnamefont {David}\ \bibnamefont {Wands}},\ }\bibfield  {title} {\enquote {\bibinfo {title} {{The Cosmological gravitational wave background from primordial density perturbations}},}\ }\href {\doibase 10.1103/PhysRevD.75.123518} {\bibfield  {journal} {\bibinfo  {journal} {Phys. Rev. D}\ }\textbf {\bibinfo {volume} {75}},\ \bibinfo {pages} {123518} (\bibinfo {year} {2007})},\ \Eprint {http://arxiv.org/abs/gr-qc/0612013} {arXiv:gr-qc/0612013} \BibitemShut {NoStop}%
\bibitem [{\citenamefont {Baumann}\ \emph {et~al.}(2007)\citenamefont {Baumann}, \citenamefont {Steinhardt}, \citenamefont {Takahashi},\ and\ \citenamefont {Ichiki}}]{Baumann:2007zm}%
  \BibitemOpen
  \bibfield  {author} {\bibinfo {author} {\bibfnamefont {Daniel}\ \bibnamefont {Baumann}}, \bibinfo {author} {\bibfnamefont {Paul~J.}\ \bibnamefont {Steinhardt}}, \bibinfo {author} {\bibfnamefont {Keitaro}\ \bibnamefont {Takahashi}}, \ and\ \bibinfo {author} {\bibfnamefont {Kiyotomo}\ \bibnamefont {Ichiki}},\ }\bibfield  {title} {\enquote {\bibinfo {title} {{Gravitational Wave Spectrum Induced by Primordial Scalar Perturbations}},}\ }\href {\doibase 10.1103/PhysRevD.76.084019} {\bibfield  {journal} {\bibinfo  {journal} {Phys. Rev. D}\ }\textbf {\bibinfo {volume} {76}},\ \bibinfo {pages} {084019} (\bibinfo {year} {2007})},\ \Eprint {http://arxiv.org/abs/hep-th/0703290} {arXiv:hep-th/0703290} \BibitemShut {NoStop}%
\bibitem [{\citenamefont {Kohri}\ and\ \citenamefont {Terada}(2018)}]{Kohri:2018awv}%
  \BibitemOpen
  \bibfield  {author} {\bibinfo {author} {\bibfnamefont {Kazunori}\ \bibnamefont {Kohri}}\ and\ \bibinfo {author} {\bibfnamefont {Takahiro}\ \bibnamefont {Terada}},\ }\bibfield  {title} {\enquote {\bibinfo {title} {{Semianalytic calculation of gravitational wave spectrum nonlinearly induced from primordial curvature perturbations}},}\ }\href {\doibase 10.1103/PhysRevD.97.123532} {\bibfield  {journal} {\bibinfo  {journal} {Phys. Rev. D}\ }\textbf {\bibinfo {volume} {97}},\ \bibinfo {pages} {123532} (\bibinfo {year} {2018})},\ \Eprint {http://arxiv.org/abs/1804.08577} {arXiv:1804.08577 [gr-qc]} \BibitemShut {NoStop}%
\bibitem [{\citenamefont {Schmitz}(2021)}]{Schmitz:2020syl}%
  \BibitemOpen
  \bibfield  {author} {\bibinfo {author} {\bibfnamefont {Kai}\ \bibnamefont {Schmitz}},\ }\bibfield  {title} {\enquote {\bibinfo {title} {{New Sensitivity Curves for Gravitational-Wave Signals from Cosmological Phase Transitions}},}\ }\href {\doibase 10.1007/JHEP01(2021)097} {\bibfield  {journal} {\bibinfo  {journal} {JHEP}\ }\textbf {\bibinfo {volume} {01}},\ \bibinfo {pages} {097} (\bibinfo {year} {2021})},\ \Eprint {http://arxiv.org/abs/2002.04615} {arXiv:2002.04615 [hep-ph]} \BibitemShut {NoStop}%
\bibitem [{\citenamefont {Yagi}\ and\ \citenamefont {Seto}(2011)}]{Yagi:2011wg}%
  \BibitemOpen
  \bibfield  {author} {\bibinfo {author} {\bibfnamefont {Kent}\ \bibnamefont {Yagi}}\ and\ \bibinfo {author} {\bibfnamefont {Naoki}\ \bibnamefont {Seto}},\ }\bibfield  {title} {\enquote {\bibinfo {title} {{Detector configuration of DECIGO/BBO and identification of cosmological neutron-star binaries}},}\ }\href {\doibase 10.1103/PhysRevD.83.044011} {\bibfield  {journal} {\bibinfo  {journal} {Phys. Rev. D}\ }\textbf {\bibinfo {volume} {83}},\ \bibinfo {pages} {044011} (\bibinfo {year} {2011})},\ \bibinfo {note} {[Erratum: Phys.Rev.D 95, 109901 (2017)]},\ \Eprint {http://arxiv.org/abs/1101.3940} {arXiv:1101.3940 [astro-ph.CO]} \BibitemShut {NoStop}%
\bibitem [{\citenamefont {Thrane}\ and\ \citenamefont {Romano}(2013)}]{Thrane:2013oya}%
  \BibitemOpen
  \bibfield  {author} {\bibinfo {author} {\bibfnamefont {Eric}\ \bibnamefont {Thrane}}\ and\ \bibinfo {author} {\bibfnamefont {Joseph~D.}\ \bibnamefont {Romano}},\ }\bibfield  {title} {\enquote {\bibinfo {title} {{Sensitivity curves for searches for gravitational-wave backgrounds}},}\ }\href {\doibase 10.1103/PhysRevD.88.124032} {\bibfield  {journal} {\bibinfo  {journal} {Phys. Rev. D}\ }\textbf {\bibinfo {volume} {88}},\ \bibinfo {pages} {124032} (\bibinfo {year} {2013})},\ \Eprint {http://arxiv.org/abs/1310.5300} {arXiv:1310.5300 [astro-ph.IM]} \BibitemShut {NoStop}%
\bibitem [{\citenamefont {Ade}\ \emph {et~al.}(2021)\citenamefont {Ade} \emph {et~al.}}]{BICEP:2021xfz}%
  \BibitemOpen
  \bibfield  {author} {\bibinfo {author} {\bibfnamefont {P.~A.~R.}\ \bibnamefont {Ade}} \emph {et~al.} (\bibinfo {collaboration} {BICEP, Keck}),\ }\bibfield  {title} {\enquote {\bibinfo {title} {{Improved Constraints on Primordial Gravitational Waves using Planck, WMAP, and BICEP/Keck Observations through the 2018 Observing Season}},}\ }\href {\doibase 10.1103/PhysRevLett.127.151301} {\bibfield  {journal} {\bibinfo  {journal} {Phys. Rev. Lett.}\ }\textbf {\bibinfo {volume} {127}},\ \bibinfo {pages} {151301} (\bibinfo {year} {2021})},\ \Eprint {http://arxiv.org/abs/2110.00483} {arXiv:2110.00483 [astro-ph.CO]} \BibitemShut {NoStop}%
\bibitem [{\citenamefont {Abazajian}\ \emph {et~al.}(2016)\citenamefont {Abazajian} \emph {et~al.}}]{CMB-S4:2016ple}%
  \BibitemOpen
  \bibfield  {author} {\bibinfo {author} {\bibfnamefont {Kevork~N.}\ \bibnamefont {Abazajian}} \emph {et~al.} (\bibinfo {collaboration} {CMB-S4}),\ }\bibfield  {title} {\enquote {\bibinfo {title} {{CMB-S4 Science Book, First Edition}},}\ }\href@noop {} {\  (\bibinfo {year} {2016})},\ \Eprint {http://arxiv.org/abs/1610.02743} {arXiv:1610.02743 [astro-ph.CO]} \BibitemShut {NoStop}%
\bibitem [{\citenamefont {Hanany}\ \emph {et~al.}(2019)\citenamefont {Hanany} \emph {et~al.}}]{NASAPICO:2019thw}%
  \BibitemOpen
  \bibfield  {author} {\bibinfo {author} {\bibfnamefont {Shaul}\ \bibnamefont {Hanany}} \emph {et~al.} (\bibinfo {collaboration} {NASA PICO}),\ }\bibfield  {title} {\enquote {\bibinfo {title} {{PICO: Probe of Inflation and Cosmic Origins}},}\ }\href@noop {} {\  (\bibinfo {year} {2019})},\ \Eprint {http://arxiv.org/abs/1902.10541} {arXiv:1902.10541 [astro-ph.IM]} \BibitemShut {NoStop}%
\bibitem [{\citenamefont {Bassett}\ \emph {et~al.}(2006)\citenamefont {Bassett}, \citenamefont {Tsujikawa},\ and\ \citenamefont {Wands}}]{Bassett:2005xm}%
  \BibitemOpen
  \bibfield  {author} {\bibinfo {author} {\bibfnamefont {Bruce~A.}\ \bibnamefont {Bassett}}, \bibinfo {author} {\bibfnamefont {Shinji}\ \bibnamefont {Tsujikawa}}, \ and\ \bibinfo {author} {\bibfnamefont {David}\ \bibnamefont {Wands}},\ }\bibfield  {title} {\enquote {\bibinfo {title} {{Inflation dynamics and reheating}},}\ }\href {\doibase 10.1103/RevModPhys.78.537} {\bibfield  {journal} {\bibinfo  {journal} {Rev. Mod. Phys.}\ }\textbf {\bibinfo {volume} {78}},\ \bibinfo {pages} {537--589} (\bibinfo {year} {2006})},\ \Eprint {http://arxiv.org/abs/astro-ph/0507632} {arXiv:astro-ph/0507632} \BibitemShut {NoStop}%
\end{thebibliography}%

\end{document}